%% file: bhc.tex
\documentclass[a4paper,fleqn,usenatbib]{mnras}

\usepackage[T1]{fontenc}
\usepackage{ae,aecompl}

\usepackage{newtxtext,newtxmath}

\usepackage{graphicx}	
\usepackage{amsmath}	
\usepackage{amssymb}	
\usepackage{siunitx, tensor, subfig, footmisc, longtable, tabu, dcolumn, booktabs}
\usepackage[referable]{threeparttablex}
\usepackage{results-ssplNs, results-adplNs}
\usepackage{results-coms}
\usepackage[inline]{enumitem}

\newcommand{\atd}{ATLAS\(^{\mathrm{3D}}\)\,}
\newcommand{\ml}[1]{\(\left(M/L\right)_{\mathrm{#1}}\)}
\DeclareMathOperator{\Msun}{~\si{M_\odot}}
\newcommand{\mi}{Model {\bf I}}{}
\newcommand{\mii}{Model {\bf II}}{}
\newcommand{\miii}{Model {\bf III}}{}
\newcommand{\mb}{Model {\bf (B)}}{}
{}

\newcommand{\fdmBValSev}{9}
\newcommand{\fdmFValSev}{13}
\newcommand{\fdmBVal}{17}
\newcommand{\fdmFVal}{19}

\newcolumntype{d}[1]{D{.}{.}{#1}}
\newcommand{\mc}[1]{\multicolumn{1}{c}{#1}}

\LetLtxMacro{\oldsqrt}{\sqrt} 


\title[Mass Profiles]{Systematic trends in total-mass profiles from dynamical models of early-type galaxies}

\author[A. Poci et al.]{
Adriano Poci$^{1}$\thanks{E-mail: adriano.poci@students.mq.edu.au (MQU)},
Michele Cappellari$^{2}$,
Richard M. McDermid$^{1,3}$
\\
$^{1}$Department of Physics and Astronomy, Macquarie University, North Ryde, NSW 2109, Australia\\
$^{2}$Sub-department of Astrophysics, Department of Physics, University of Oxford, Denys Wilkinson Building, Keble Road, Oxford OX1 3RH, UK\\
$^{3}$Australian Astronomical Observatory, PO Box 915, Sydney NSW 1670, Australia
}

\date{Accepted XXX. Received YYY; in original form 2016 August 21}

\pubyear{2016}

\begin{document}
\label{firstpage}
\pagerange{\pageref{firstpage}--\pageref{lastpage}}
\maketitle

\begin{abstract}
We study trends in the slope of the total mass profiles and dark matter fractions within the central half-light radius of 258 early-type galaxies, using data from the volume-limited \atd survey. We use three distinct sets of dynamical models, which vary in their assumptions and also allow for spatial variations in the stellar mass-to-light ratio, to test the robustness of our results. We confirm that the slopes of the total mass profiles are approximately isothermal, and investigate how the total-mass slope depends on various galactic properties. The most statistically-significant correlations we find are a function of either surface density, \(\Sigma_e\), or velocity dispersion, \(\sigma_e\). However there is evidence for a break in the latter relation, with a nearly universal logarithmic slope above \(\log_{10}[\sigma_e/(\si{km~s^{-1}})]\sim 2.1\) and a steeper trend below this value. For the 142 galaxies above that critical \(\sigma_e\) value, the total mass-density logarithmic slopes have a mean value \(\left\langle\gamma^\prime\right\rangle =  \slTVal \pm \slTValErr\) (\(1\sigma\) error) with an observed rms scatter of only \(\sigma_{\gamma^\prime}=\kerWidTVal \pm \kerWidTValErr\). Considering the observational errors, we estimate an intrinsic scatter of \(\sigma_{\gamma^\prime}^\mathrm{intr} \approx 0.15\). These values are broadly consistent with those found by strong lensing studies at similar radii and agree, within the tight errors, with values recently found at much larger radii via stellar dynamics or HI rotation curves (using significantly smaller samples than this work).
\end{abstract}




\section{Introduction}\label{sec:intr}
The current paradigm of galaxy formation dictates that the baryonic matter that forms the galaxies we observe is embedded in large haloes of dark matter \citep{white78}. It is believed that dark matter, while unseen by conventional observations, is the dominant form of mass in the Universe \citep{blumenthal84}. Meanwhile, galactic observables are derived almost exclusively from observations of the baryonic matter. Yet, as crucial to galaxy formation as these two components are, the interplay between dark and baryonic matter on galactic scales has so far evaded a detailed explanation. Unfortunately the empirical separation between the luminous and dark matter suffers from observational difficulties and model degeneracies, unless very spatially extended kinematic information is available. However, what is robust to measure is the combination of both components, namely the total mass distribution. For this reason, the total mass has emerged recently as an important empirical test for galaxy formation theories.\par
Since the first convincing evidence for dark matter in galaxies \citep{rubin70}, rotation curves have been a critical tool for studying late-type (spiral) galaxies. The rotational velocities are set by the gravitational potential in which the galaxy resides, and thus trace the total mass. Measurements of these rotational velocities revealed that they remain approximately constant out to large galactocentric radii, rather than declining in velocity as would be expected for standard Keplerian orbits. Known as the disk-halo conspiracy \citep{bahcall85, albada86}, this observation is interesting on account of the broad range of masses that have been studied, using a variety of observational techniques. Subsequently, attention has since turned to early-type galaxies (ETGs). The Sloan Lens Advanced Camera for Surveys (SLACS) project \citep{bolton06} found that the total mass-density profile slopes appear to vary little across a sample of ETGs with stellar mass \(M_\ast\ga 10^{11} \Msun\) \citep{koopmans09, auger10}. Isotropic mass models were constructed for 73 confirmed strong lenses with median Einstein radius of \(R_e/2\). The mean slope of the power-law density profile \(\rho_{\mathrm{tot}}(r) \propto r^{-\gamma^{\prime}}\) was found to be \(\left\langle\gamma^\prime\right\rangle = 2.078 \pm 0.027\) with an intrinsic scatter of just \(\sigma_{\gamma^\prime}^\mathrm{intr} = 0.16 \pm 0.02\) \citep{auger10}. In the case of ETGs, the non-isothermal stellar and dark components must `conspire' in such a way as to consistently produce a nearly-isothermal total profile (where an isothermal profile is \(\rho(r)\propto r^{-2}\)), across the entire sample studied by \cite{auger10}.\par
\cite{cappellari15} constructed anisotropic dynamical models of the total mass, for a sample of 14 fast-rotator ETGs. Unlike previous work, they were able to probe the outer regions of their objects (to a median radius of \(4~R_e\)), where dark matter is believed to dominate. They find comparable results, reporting a mean slope and scatter for the total mass-density profiles of \(\left\langle\gamma^\prime\right\rangle = 2.19 \pm 0.03\) and \(\sigma_{\gamma^\prime} = 0.11\), respectively. Using observations of \nuclide{HI}, \cite{serra16} were able to model a sample of 16 fast-rotator ETGs out to a mean radius of \(6~R_e\), and independently confirm, with remarkable accuracy, both the slope and scatter previously found, reporting \(\left\langle\gamma^\prime\right\rangle = 2.18 \pm 0.03\) and \(\sigma_{\gamma^\prime} = 0.11\). These studies provide very strong evidence for the existence of the bulge-halo conspiracy in ETGs.\par
Exploring possible relations of the total mass-density profile slope with galactic properties, within \(R\approx R_e/2\), \cite{dutton14} modelled various galaxy-formation scenarios within the \(\Lambda\)-Cold-Dark-Matter paradigm. They found that the observations of the bulge-halo conspiracy are best reproduced by models that assume an uncontracted Navarro, Frenk, and White \citep[NFW;][]{navarro96} dark matter halo. They also presented model predictions for the correlations of the total mass-density profile slope (within \(1~R_e\)) with stellar surface density, effective radius, stellar mass, and velocity dispersion, and found correlations with all of these parameters (albeit with varying degrees of significance). Using kinematic modelling, \cite{cappellari15} and \cite{serra16} investigated possible correlations of the slope out to \(\sim4~R_e\) with effective radius and velocity dispersion, and found no significant trends. However, their samples spanned a limited range of mass and were restricted to \(\sigma_e \ga 100~\si{km s^{-1}}\), and this may be the reason for the lack of trends. In fact, when using the models of \cite{cappellari13a} and restricting the analysis to the range covered by the \atd observation (\(1~R_e\)), \citet[][Fig. 22c]{cappellari16} found that, below the critical mass \(M_{\rm crit}\la 2\times10^{11} \Msun\), the total profile slopes follow lines of nearly constant velocity dispersion. This agrees with the indications from the gradients of \(\sigma(R)\) presented in \cite{cappellari13b}.\par
Numerical simulations have also produced similar results. \cite{remus13} concluded that galactic evolution via mergers drives the total-mass towards a slope of \(\sim -2\). Furthermore, they claim that once a galaxy has evolved such that its total mass is described by a power-law with slope of \(\sim -2\), subsequent mergers do not cause departures from this value, indicating that this slope does indeed have unique properties in the context of galaxy formation and evolution.\par
In this work, we present the results of our own modelling on the correlations of total mass-density profile slope with galactic observables, as well as the mean and scatter of both the total and stellar-only slope distributions. We use stellar kinematics for 258 ETGs from the \atd survey \citep{cappellari11}, which extends to a median radius of \(0.9~R_e\) \citep{emsellem11}. This makes our spatial coverage comparable to that of the SLACS survey. However, the two samples are quite complementary, with \atd being essentially mass-selected, while SLACS is nearly \(\sigma\)-selected \citep[see][for a discussion]{posacki15}. In addition to kinematic models, we construct detailed star formation histories from spectroscopic data in order to account for effects of spatial variations in the stellar populations. These factors produce robust, homogeneous, and general models of a large sample of ETGs, which in turn place strong constraints on the existence of a `bulge-halo conspiracy' and any correlations with galactic observables.\par
This paper is laid out as follows. \autoref{sec:data-model} describes the data and modelling processes in detail, including the construction of the stellar population models, as well as the assumptions behind our various models. The results are presented in \autoref{sec:results}, which compares all models to previous work. A discussion and interpretation of these results are given here also. Our conclusions are given in \autoref{sec:concl}.

\section{Data \& Model Definitions}\label{sec:data-model}
\subsection{Kinematic Data}
The kinematic data for this work comes from the \atd survey \citep{cappellari11}. \atd studied 260 ETGs with the SAURON \citep{bacon01} Integral-Field Unit (IFU). Objects were selected to have \(D < 42~\si{Mpc}\) and \(M_K < -21.5~\si{mag}\), with the resulting sample containing stellar masses of \(M_{\star} \gtrsim 6\times 10^9~\si{M_\odot}\). The IFU data\footnote{\label{fn:atlas}Available from http://purl.org/atlas3d} provides both the kinematic and spectral data necessary for the modelling conducted in this work. For a description of the extraction of the stellar kinematics, see \cite{cappellari11}.

\subsection{Spectral Models}\label{ssec:specmod}
In order to account for spatial variations in the stellar populations, we compute the radial profile of the stellar mass-to-light ratio (\(M/L\)) for each galaxy. A Salpeter IMF is assumed for this derivation, and so the result is denoted \ml{Salp}. The data are Voronoi-binned \citep{cappellari03}, with a minimum \(S/N\) of \(80\) per bin. Each bin is fit with a linear combination of single stellar population model spectra using the penalised pixel fitting (pPXF) method of \cite{cappellari04}, and upgraded as described in \cite{cappellari17}. We use 66 single stellar population (SSP) templates taken from the MIUSCAT model library \citep{vazdekis12}. Our templates span a grid of 11 ages, with \(t\) \(=\) \([\)\(0.501,\) \(1.000,\) \(1.995,\) \(3.162,\) \(3.981,\) \(5.012,\) \(6.310,\) \(7.943,\) \(10.000,\) \(12.589,\) \(14.125\)\(]~\si{Gyr}\) (chosen for computational speed), and 6 metallicities, with \([Z/H]\) \(=\) \([\)\(-1.71,\) \(-1.31,\) \(-0.71,\) \(-0.40,\) \(0.00,\) \(0.22\)\(]\) 
(equivalent to \(Z=0.0004 - 0.03\)), and assuming a unimodal IMF of slope 1.30 \citep[equivalent to][]{salpeter55}. In the pPXF fit, we employ linear regularization to suppress noise in the weights distribution and produce smooth solutions, whereby the weights of the fit vary gradually in age and metallicity \citep[as done in][]{onodera12, cappellari12, mcdermid15, posacki15, morelli13, morelli15}. We compute this fit at every bin, and use the resulting star formation histories to produce a spatially-resolved map of \ml{Salp}. These values\footref{fn:atlas} are then parametrised on thin elliptical annuli in order to construct the \ml{Salp}\((R)\) profile. For each of the \(j\) Gaussians of the stellar mass model (see \autoref{ssec:adm}), a corresponding annulus is constructed that extends out to \(\sigma_j\), with ellipticity \(q_j\). This process hence relaxes the assumption of a spatially-constant \ml{Salp} from the initial \atd study \citep{cappellari13a}. We note that that work used a different grid of ages in their SSP templates, with 26 logarithmically-spaced ages from \(1-17.782~\si{Gyr}\).

\subsection{Axisymmetric Dynamical Models}\label{ssec:adm}
All kinematic models conducted for this work use the multi-Gaussian Expansion (MGE) technique \citep{monnet92, emsellem94, cappellari02}, in conjunction with Jeans Anisotropic Modelling (JAM) \citep{cappellari08}. The MGE provides an analytic description of the observed surface brightness, by fitting a series of Gaussians to the isophotes. The MGEs for this work were computed from the Sloan Digital Sky Survey \citep[SDSS;][]{york00} and Isaac Newton Telescope {\it r}-band (AB mag) photometry, and were published in \cite{scott13}\footref{fn:atlas}. Thus, all \(L\) and \(M/L\) measurements in this work are implicitly {\it r}-band measurements. The JAM model then predicts the second moment of the velocity distribution based on this MGE, assuming the galaxy is axisymmetric.\par
We run and compare three distinct models in this work, which are described in detail below.
\begin{description}
    \item[{\it Model I}] This model assumes the \emph{total} mass-density profile is a spherical double power-law of the form
\begin{equation}\label{eq:spl}
    \rho_{\mathrm{tot}}\left(r\right) = \rho_s \left( \frac{r}{r_s} \right)^{\gamma} \left( \frac{1}{2} + \frac{1}{2} \frac{r}{r_s} \right)^{-\gamma - 3}
\end{equation}
where \(r_s\) is the break radius and \(\rho_s\) is the density at the break radius. This is a special case of the Nuker Law \citep{lauer95}, with \(\alpha=1\). However, since we fix \(r_s\) at \(20~\si{kpc}\) \citep[as done in][]{cappellari13a}, which is much larger than the field-of-view of the observations, this model approximates a single power-law in the region where we have data. Contrary to standard practise, the gravitational potential is not computed as the co-addition of the stellar mass and an assumed parametrisation for the dark matter halo. Instead, the gravitational potential is computed from the assumed spherical parametrisation of the total density, given in \autoref{eq:spl}. The stellar MGE is simply used by the JAM model to describe the distribution of the stellar tracer, which is needed to compute the dynamical observables. This non-standard approach was also used in model (e) of \cite{mitzkus17}. Of course, the total density will include a contribution from the stars and the dark matter, and account for possible variations in the stellar \(M/L\). However, in our approach, no assumptions are needed about how these different components contribute to the total density. Only \emph{after} the fit, this total mass profile is decomposed into its corresponding stellar and dark matter constituents, as described in \autoref{ssec:decomp}. This model has 4 free parameters;
\begin{enumerate*}[label=(\roman*)]
    \item orbital anisotropy, \(\beta_z\equiv 1 - \sigma_z^2/\sigma_R^2\), for velocity dispersion along the symmetry axis and cylindrical radius, \(\sigma_z\) and \(\sigma_R\), respectively.
    \item inclination, \(i\)
    \item density at the break radius, \(\rho_s\)
    \item inner slope of the total mass profile, \(\gamma\)
\end{enumerate*}.
    \item[{\it Model II}] For this model, the \emph{total} mass-density profile is assumed to be an axisymmetric oblate spheroid given by
\begin{equation}\label{eq:dpl}
    \rho_{\mathrm{tot}}\left(m\right) = \rho_s \left( \frac{m}{m_s} \right)^{\gamma} \left( \frac{1}{2} + \frac{1}{2} \frac{m}{m_s} \right)^{-\gamma + \nu}
\end{equation}
where \(m = \sqrt{R^2 + z^2/q^2}\) is the elliptical radius and \((R,z)\) are cylindrical coordinates, now \(\gamma\) is the inner slope, and \(\nu\) is the outer slope. In this model also, the stars still act only as a tracer and are described by the MGEs of \cite{scott13}. Here, we do not fix \(r_s\). It is instead added as an additional free parameter, with generous bounds to avoid prohibiting any good solutions. As a result, this model has 7 free parameters;
\begin{enumerate*}
    \item orbital anisotropy, \(\beta_z\)
    \item inclination, \(i\)
    \item density at the break radius, \(\rho_s\)
    \item inner slope of the total mass profile, \(\gamma\)
    \item outer slope of the total mass profile, \(\nu\)
    \item break radius, \(r_s\)
    \item axis ratio of the oblate spheroid for the total mass component, \(q\)
\end{enumerate*}.
    \item[{\it Model III}] In this case we simply compute the total mass slopes from the published \mb\ of \cite{cappellari13a} without re-fitting the model. Unlike the above two, this model does not explicitly parameterise the total mass. It instead takes the more conventional approach of fitting the measured stellar and assumed dark matter components directly to the data. The total mass of this model is described as the co-addition of a spherical NFW halo and the MGE stellar density for an assumed constant \(M/L\) from \cite{scott13}. We thus use \miii\ as a baseline to compare against the total mass approach. This model had 4 free parameters;
\begin{enumerate*}
    \item orbital anisotropy, \(\beta_z\)
    \item inclination, \(i\)
    \item stellar mass-to-light, \ml{stars}
    \item the dark matter halo virial mass, \(M_{200}\)
\end{enumerate*}.
\end{description}
The parameters are optimised in a Bayesian framework, using Markov Chain Monte Carlo (MCMC) inference techniques \citep{hastings70}. We use \textsc{emcee} \citep{mackey13}, a Python implementation of the \cite{goodman10} Affine-Invariant MCMC Ensemble sampler, to estimate the posterior distributions of our free parameters. Flat priors were assumed for all parameters within their respective bounds. The bounds for each parameter are illustrated by the extents of the corresponding panels in \autoref{img:mcmc}. Some parameters have hard bounds based on physical reasoning. Inclination can not exceed \(90\si{\degree}\), while the lower limit is determined by the flattest Gaussian of the MGE model. The upper bound for both the inner and outer slopes is set to \(0\), since it is unreasonable to expect the total density to increase with radius. The axis ratio is defined on the interval \([0,1]\). All other bounds were chosen simply to give the walkers sufficient freedom when probing the space.\par
We then characterise the distributions of total- and stellar-mass density profile slopes from our models. The slope we use for all models is the mean logarithmic slope, computed by definition as
\begin{equation}\label{eq:logslope}
\begin{aligned}
    \gamma^\prime &= \Delta\log\rho_{\rm tot} \big/ \Delta\log r\\
    &= \frac{\log\left[\rho\left(R_{\rm out}\right) / \rho\left(R_{\rm in}\right)\right]}{\log_{10}\left[R_{\rm out} / R_{\rm in}\right]}
\end{aligned}
\end{equation}
where
\begin{equation}\label{eq:bounds}
\begin{aligned}
    R_{\rm in} &= {\rm max}\left(R_e/10, 2\si{\arcsecond}\right)\\
    R_{\rm out} &= {\rm min}\left(R_e, r_{\rm max}\right)
\end{aligned}
\end{equation}.
We enforce the conditions in \autoref{eq:bounds} to ensure that we are using only constrained data to compute the slopes. The inner bound, \(R_{\rm in}\), is set so as to reduce the impact of the flattening of the stellar MGE on the computation of the slope. This flattening is an effect of seeing on the SDSS images from which the MGEs were derived, rather than a physical characteristic of the galaxy. The outer bound, \(R_{\rm out}\), is set such that for the majority of the sample we use \(R_e\) as is typical for galactic property derivations. However, in the few cases within our sample where \(R_e > R_{\rm max}\), this condition ensures that no extrapolated data is used to compute the slope.\par
Since \(r_S\) is fixed at \(20~\si{kpc}\) in \mi, \(\gamma^\prime\) essentially coincides with the parameter \(\gamma\) in \autoref{eq:spl}, as directly optimised in MCMC. Conversely, given the freedom in \(r_S\) and \(\nu\), this may not be the case in \mii.

\subsection{Decomposing Dark and Luminous Mass}\label{ssec:decomp}
In most previous work \citep[for example, see][]{weijmans08, forestell10, morganti13, cappellari13a, cappellari15}, one typically assumes a certain parametrisation for both the stars and the dark matter. In that case, the free parameters of the models consist of the \(M/L\) of the axisymmetric/triaxial stellar distribution, as well as the parameters of the generally-spherical dark matter halo.\par
Here we use a different approach, which is motivated by our desire to reduce the number of assumptions made in the derivation of the total mass-density profiles, which is the main focus of this work. To achieve this, we parametrize the total mass distribution directly and we treat the stars as tracer particles in the adopted gravitational potential. In this way, our total mass profiles are only dependent on our rather general assumed parametrisations given in Eqs.~\ref{eq:spl} and \ref{eq:dpl}, while it is completely independent of any assumption about the luminous and dark matter.\par
Once the total density has been obtained, to make inferences about the luminous and dark matter separately, we need to introduce assumptions about their parametrizations. While we do not claim that this approach is superior to fitting the kinematics assuming a stellar and dark matter halo as in \miii, the approach of Models {\bf I} and {\bf II} is distinctly different from \miii, which allows us to assess the robustness of decomposing the stellar and dark matter profiles. For the dark component, we consider the NFW profile
\begin{equation}\label{eq:dm}
    \rho_{\mathrm{DM}}\left(r\right) = \rho_{\mathrm{DM},s} \left( \frac{r}{r_s} \right)^{-1} \left( \frac{1}{2} + \frac{1}{2} \frac{r}{r_s} \right)^{-2}
\end{equation}
where \(r_s = 20~\si{kpc}\) for the dark matter profile in Models {\bf I} and {\bf II}. The stellar mass is computed by multiplying the stellar luminosity by the \(M/L\) inferred from the stellar population via full spectrum fitting (see \autoref{ssec:specmod}). Specifically, we take the MGE from \cite{scott13} and multiply the peak surface brightness \(\Sigma_j\) of the \(j\)-th Gaussian (with dispersion \(\sigma_j\)) by the projected radial \ml{Salp} profile, measured at radius \(R=\sigma_j\). In doing this, we account for differences in the stellar mass that arise from variations in age and metallicity across the galaxy. After this multiplication, the MGE provides a description of the stellar mass surface density, for the assumed Salpeter IMF. This MGE is then deprojected using Eqs.~(10) and (14) of \cite{cappellari13a}, to obtain the intrinsic axisymmetric stellar density. The spherically-averaged stellar density profile is computed from the MGE using Note 11 of \cite{cappellari15}:
\begin{equation}\label{eq:mge}
    \rho_{\mathrm{star}}\left(r\right) = \alpha_{\mathrm{IMF}}\sum\limits_{j=1}^{N} \frac{M_j \exp\left( -\frac{r^2}{2 \sigma_j^2}\right) \mathrm{erf}\left( \frac{r\sqrt{1 - q_j^2}}{q_j \sigma_j \sqrt{2}}\right)}{4 \pi \sigma_j^2 r \sqrt{1 - q_j^2}}
\end{equation}
where \(M_j\), \(\sigma_j\), and \(q_j\) are the deprojected mass, dispersion, and axis ratio, respectively, of the \(j\)-th Gaussian of the MGE model. \(\alpha_{\rm IMF}\) is a global scaling constant. As described in \autoref{ssec:specmod}, the stellar MGE is computed under the assumption of a constant Salpeter IMF. For this reason, any deviation of \(\alpha_{\rm IMF}\) from unity can be interpreted as a deviation of the IMF mass normalisation from Salpeter.\par
\begin{figure*}
    \centerline{
    \includegraphics[resolution=600, width=0.9\textwidth]{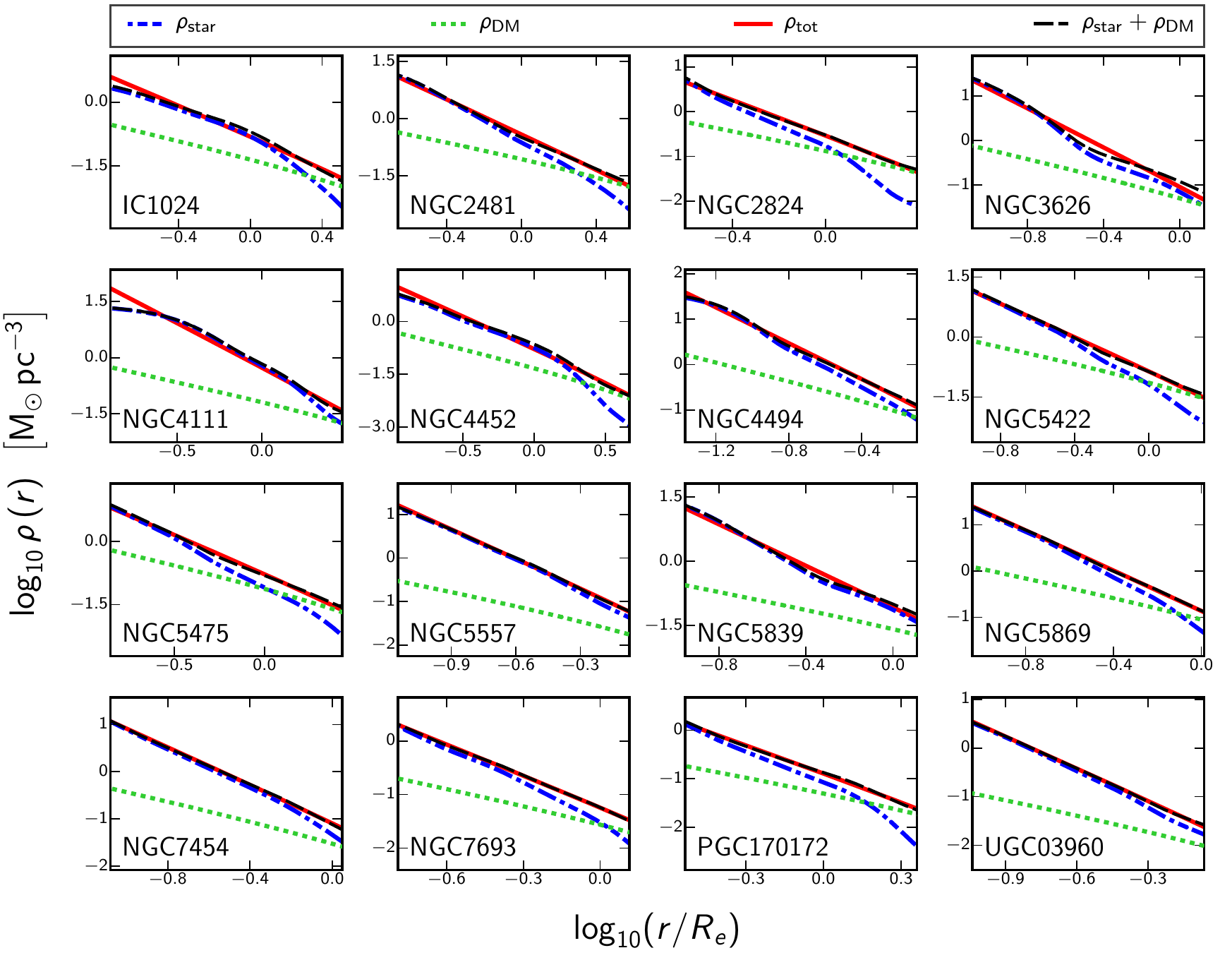}}
    \caption{The decomposition of the total mass-density profile into the stellar and dark matter components, for 16 galaxies from the \atd sample. The profiles are fitted only in the region \(2\si{\arcsecond}\leq r \leq r_{\mathrm{max}}\), which is set as the range of each \(x\)-axis, where \(r_{\mathrm{max}}\) is the outer-most elliptical coordinate in the data cube.}
    \label{img:decomp}
\end{figure*}

The only free parameters at this stage are \(\alpha_{\rm IMF}\) and \(\rho_{\mathrm{DM},s}\). Thus, we set \(\rho_{\mathrm{tot}} = \rho_{\mathrm{star}} + \rho_{\mathrm{DM}}\), and optimise these remaining free parameters in an independent run of MCMC. The final result of this optimisation for 16 representative galaxies is illustrated in \autoref{img:decomp}.
These profiles are fitted only in the region \(2\si{\arcsecond}\leq r \leq r_{\mathrm{max}}\). The central \(2\si{\arcsecond}\) is excluded from the JAM model to avoid any bias to the stellar velocities due to the presence of a supermassive black hole \citep[however, the mass of the black hole is accounted for according to the \(M_{\bullet}-\sigma\) relation of][]{gebhardt00}. Thus this region should not influence subsequent fits. Furthermore, given that a small subset of the \atd sample have \(r_{\mathrm{max}} < R_e\) (where \(r_{\mathrm{max}}\) is the largest elliptical coordinate in the kinematic data), we avoid extrapolation by using \(r_{\mathrm{max}}\), ensuring we fit only where we have kinematic data.\par
Once we have optimised all parameters, we compute the dark matter fractions and slopes for the objects in our sample. The fraction of dark matter is calculated as
\begin{equation}\label{eq:dmf}
    f_{\rm DM} (r = R_e) = \frac{M_{\rm DM}(r=R_e)}{M_{\rm tot}(r=R_e)}
\end{equation}
where \(M (r=R_e)\) is the mass enclosed in a sphere of radius \(R_e\). Specifically, \(M_{\rm tot}\) here corresponds to the integral of the total mass-density profile, \(\rho_{\rm tot}\), which is fit directly to the kinematics. \(M_{\rm tot}\) and \(M_{\mathrm{DM}}\) are analytically integrated from the total- and dark-mass MGEs, respectively (which characterise the surface {\em density}, instead of the surface brightness as described in \autoref{ssec:adm}) up to a radius of \(1 R_e\) using Eq. 1 of \cite{janz16}:
\begin{equation}
\begin{aligned}
    &M(r) = \sum\limits_{j} M_j \Big\{ {\rm erf}\left[r \big /\left(\sqrt{2} q_j \sigma_j\right) \right]\\
&\phantom{M(r) = }-\frac{{\rm exp}\left[-r^2 \big / \left(2 \sigma_j^2\right)\right]{\rm erf} \left[r \sqrt{1-q_j^2} \big/ \left(\sqrt{2} q_j \sigma_j\right)\right]}{\sqrt{1 - q_j^2} }\Big\}
\end{aligned}
\end{equation}

\section{Results \& Discussion}\label{sec:results}

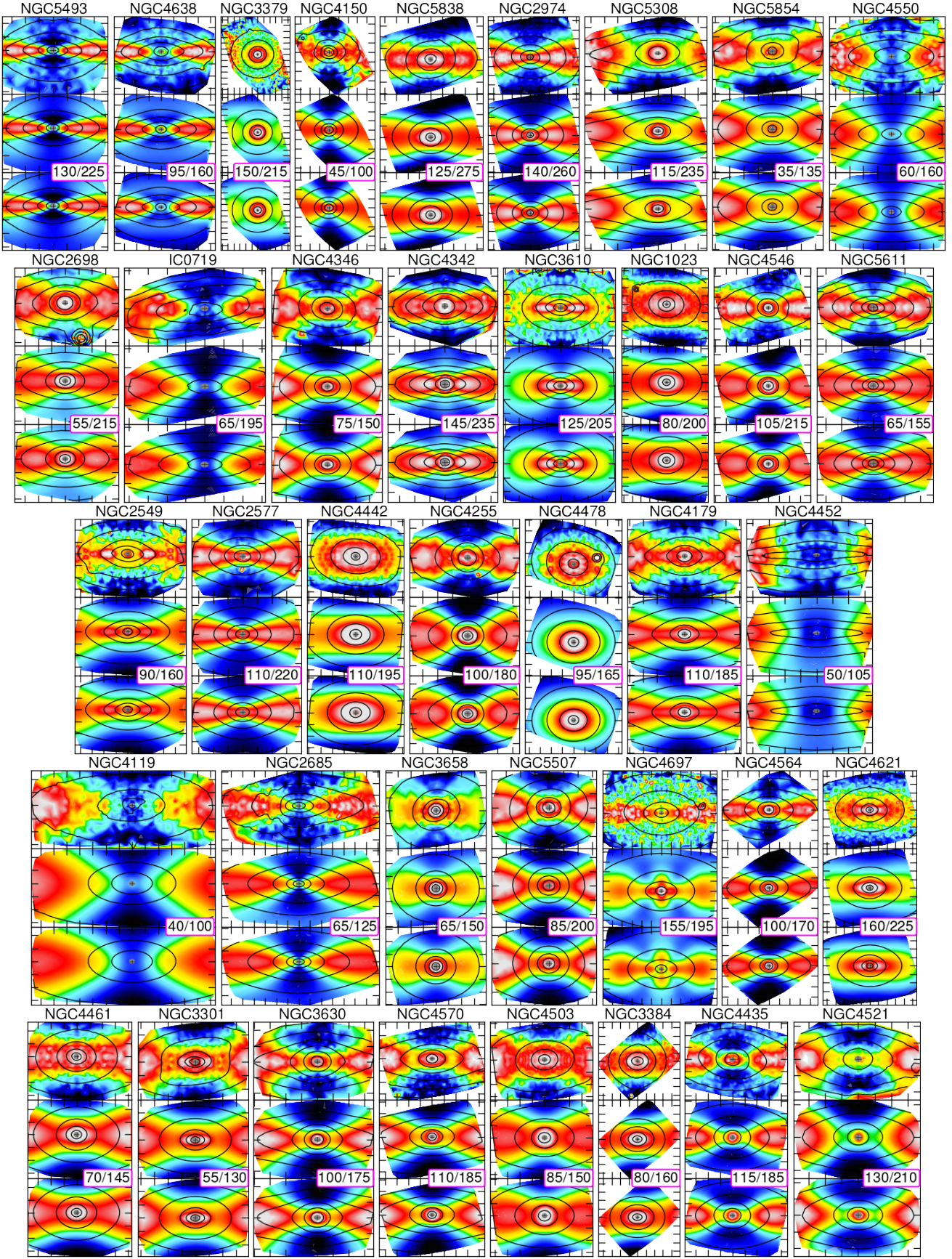
\begin{figure*}
    \input{thumbs.tex}
    \caption{In each individual plot we show the second velocity moment, \(V_{\rm RMS} = \sqrt{V^2 + \sigma^2}\), as measured by the \atd survey (top panel), as predicted by \mi\ (middle panel), and as predicted by \mii\ (bottom panel) for a subset of the sample used in this work. The inset numbers are the limits of the colour scale (\(\mathrm{min}/\mathrm{max}\), in \(\si{km~s^{-1}}\)). The tick marks are in \(5\si{\arcsecond}\) increments. The central \(2\si{\arcsecond}\) is masked with a grey ellipse, as this region is not included in the fit.}
    \label{img:jam}
\end{figure*}

\begin{figure*}
    \input{corn-thumbs.tex}
    \caption{The MCMC results for 6 objects from the sample. The approximations by MCMC to the posterior distributions, marginalized over one parameter, are shown in the diagonal panels, while the off-diagonal panels show the posterior distribution marginalized over every pair of parameters. The points are coloured according to their likelihood, with white being the most likely. The extents of each panel illustrates the parameter limits imposed on MCMC, with the tick marks in increments of; \(0.1\) for \(\beta_z\); \(5\si{\degree}\) for \(i\); \(1\) for \(\log_{10}\left(\rho_s\right)\), \(\gamma\), and \(\nu\); \(5~\si{kpc}\) for \(r_s\); and \(0.1\) for \(q\). The maps show the \(V_{\mathrm{RMS}}\) for the data (top panel) and \mii\ (bottom panel), with the limits of the colour scale inset between them.}
    \label{img:mcmc}
\end{figure*}
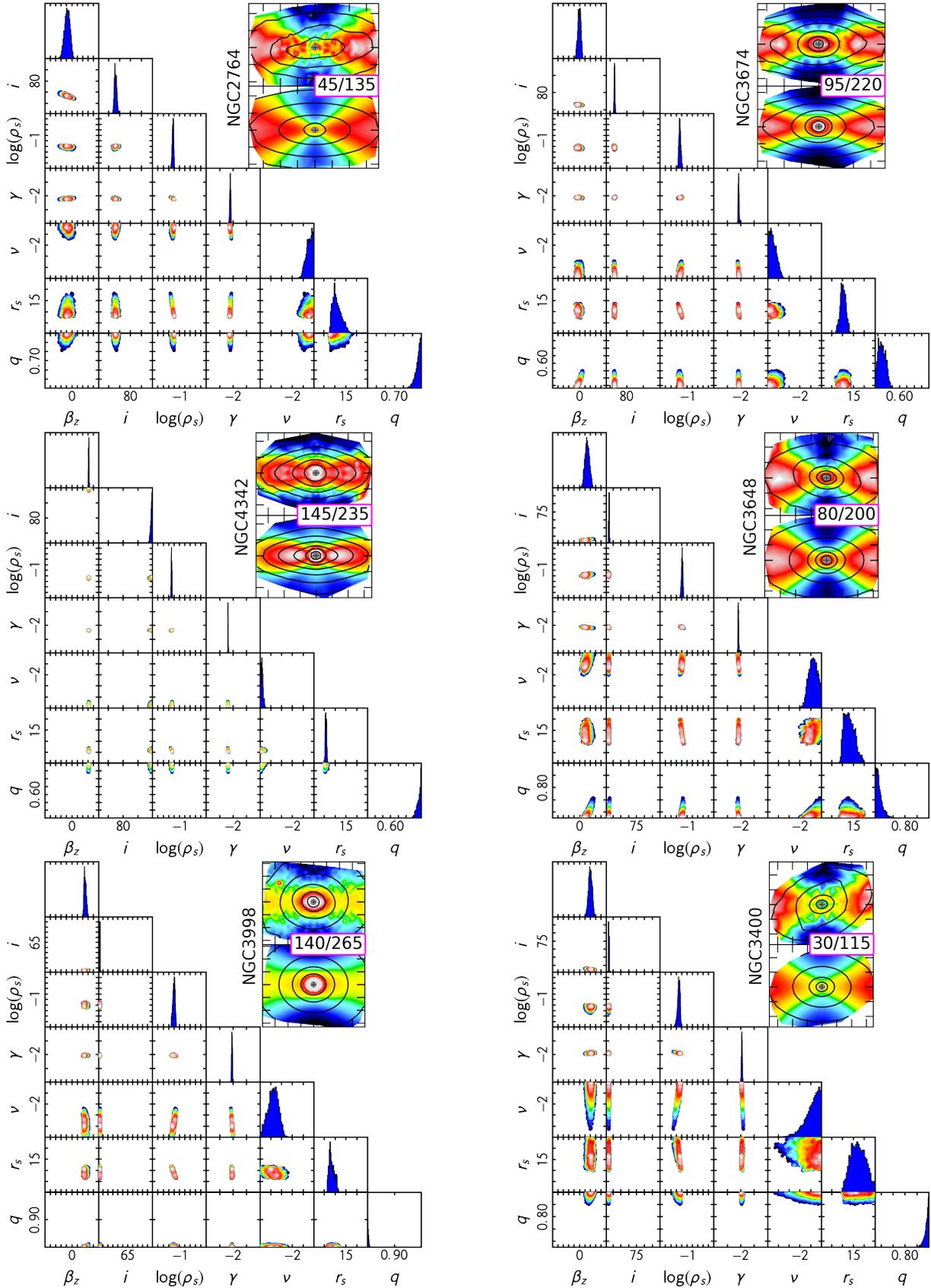

Presented here are the results of our kinematic modelling. \autoref{img:jam} presents the JAM models for a subset of the \atd sample, selected to illustrate the range of kinematic structures probed by our models.
For an in-depth discussion of the kinematics structures present in the \atd sample, see \cite{krajnovic11}. Similarly, the results of the MCMC process for a subset of the sample are given in \autoref{img:mcmc}.

\subsection{Dark Matter Content}

\begin{figure}
    \includegraphics[resolution=600, width=1\columnwidth]{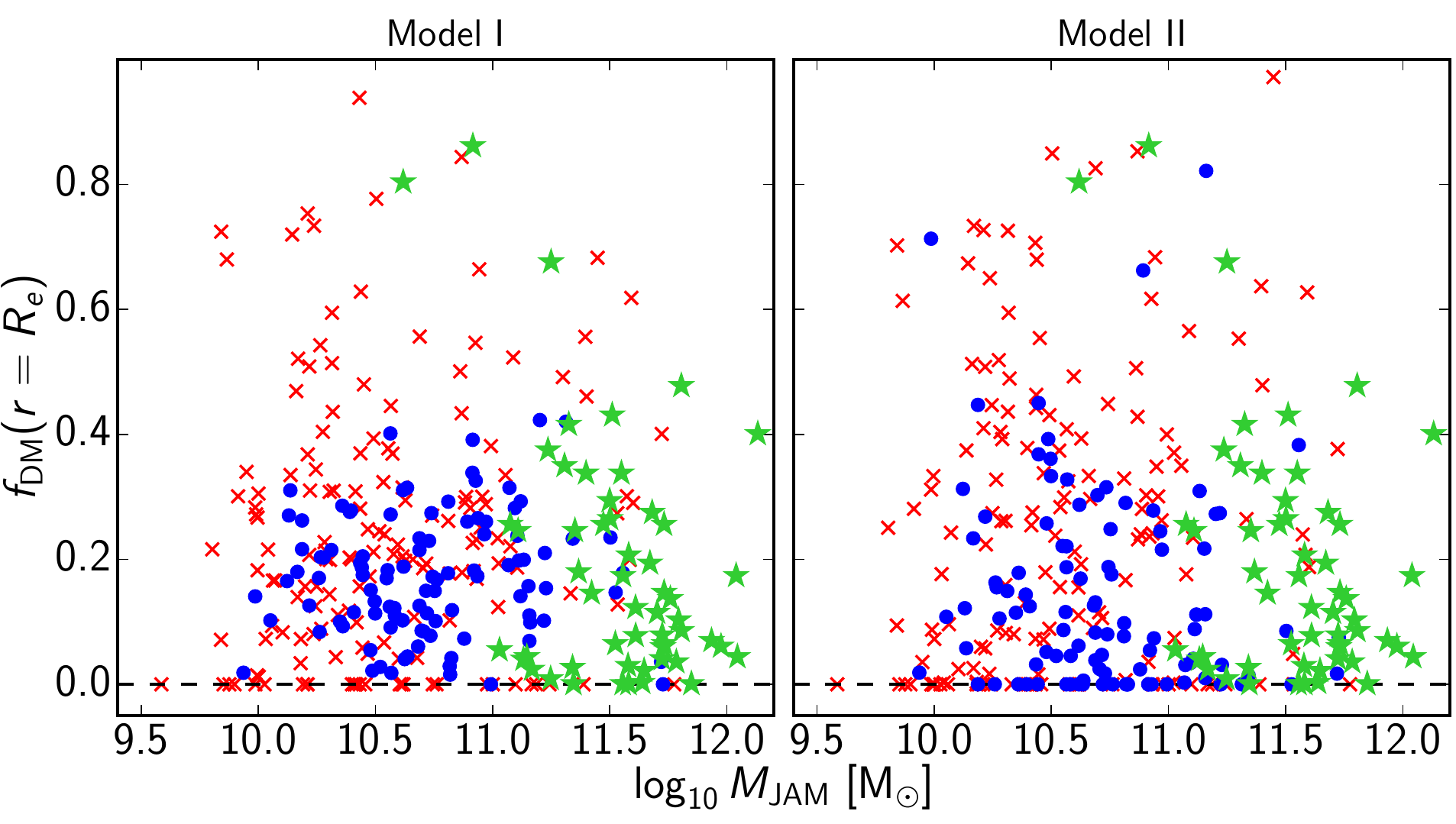}
    \caption{The fraction of dark matter within \(1~R_e\) for all the galaxies in our sample, as calculated by \mi\ (left panel) and \mii\ (right panel). The corresponding figure for \miii\ was published as the top panel of Fig. 10 in \protect\cite{cappellari13a}. The galaxies with data quality \(\leq 1\) are represented by red crosses, which were flagged for having strong dust, poor kinematic data, or other features that are detrimental to the modelling \protect\citep[see Table 1,][]{cappellari13a}. The green stars are SLACS data, as published by \protect\cite{posacki15}.}
    \label{img:dmf}
\end{figure}

The inferred dark matter fractions of the full \atd sample for Models {\bf I} and {\bf II} are shown in \autoref{img:dmf}, and recorded in \autoref{tab:data}. We extend our sample with data from \cite{posacki15}, who conducted a detailed lensing/dynamics analysis on 55 of the SLACS lenses. This data extends our sample to higher velocity dispersions.
Comparisons are presented in \autoref{tab:fdm}.
\begin{table}
    \caption{The dark matter fractions.}
    \label{tab:fdm}
    \begin{center}
    \(\begin{tabu}{cccc}
        \toprule
        f_{\rm DM} [\%] & f_{\rm DM} [\%] & N & \text{Reference}\\
        {\rm Best} & {\rm Sample} & {\rm Sample} & \\\midrule
        \fdmBVal & \fdmFVal & 258 & \text{\mi, SD}\\
        \fdmBValSev & \fdmFValSev & 258 & \text{\mii, SD}\\
        9 & 13 & 258 & \text{\miii, SD}\\
        - & 23 & 59 & \text{\protect\cite{auger10}, L}\\
        - & 14 & 55 & \text{\protect\cite{posacki15}, SD/L}\\
        \bottomrule
    \end{tabu}\)
    \end{center}
    \begin{tablenotes}[para, flushleft]
        \note The `best' and `sample' columns are for the best models \citep[with data quality \(>\) 1; see Table 1,][]{cappellari13a} and the full sample, respectively. \protect\cite{auger10} provide tabulated data on the 59 objects out of their 73 with reliable kinematics. The value for \protect\cite{posacki15} was computed by adopting the upper limits where necessary from Table 2 of that work. The type of observation used is provided with the reference, where SD is stellar dynamics, and L is lensing.
    \end{tablenotes}
\end{table} The agreement of our models with previous work is interesting on account of the significant differences between the observational techniques and sample size.\par
In the decomposition of the total mass, we attribute any difference between the stellar and total-mass components to dark matter. The resulting dark matter fractions are therefore dependent on the assumed stellar \(M/L\) profile. As mentioned above, we take into account the spatial variations in age and metallicity (the star formation history) by computing full spectral fits to each spatially-binned spaxel in the SAURON data, which in turn are converted into \(M/L\) values assuming a constant IMF (\autoref{ssec:specmod}). We note that, in addition to the overall rescaling of the stellar mass normalisation encapsulated by the parameter \(\alpha_{\mathrm{IMF}}\) in \autoref{eq:mge}, the precise shape of the stellar \(M/L\) profile for a given star formation history will depend to some extent on the choice of IMF, as the response of the stellar \(M/L\) to different age and metallicity distributions depends on the ratio of (short lived) high-mass to (long lived) low-mass stars. Thus, while the IMF itself does not vary in our work, a different choice of fiducial IMF would result in different stellar \(M/L\) profiles, with subsequently different stellar slopes and dark matter fractions. To test the impact of this, we re-computed our dark matter fractions for 12 unimodal IMFs, with slopes ranging from \(0.30\) to \(3.30\), as well as a broken power-law Kroupa-like \citep{kroupa01} IMF. We are able to exclude unimodal slopes of \(2.50\) or steeper, as they result in stellar masses in excess of the total dynamical mass for all galaxies. Considering only the IMFs with physically-permitted unimodal slopes or Kroupa-like shape, the dark matter fractions of some galaxies with relatively strong population gradients could change by up to \(0.10\), while the full-sample median dark matter fraction varied by at most \(\sim 0.04\) in absolute terms.\par
Some galaxies in the sample are affected considerably by dust, and so more sophisticated dust correction algorithms may improve the mass model derived from the photometry. A more accurate mass model would improve the accuracy of the stellar profile, and subsequently of the dark matter fractions also. Moreover, dynamics from infrared spectroscopy would further alleviate issues with dust \citep[for example, see][]{williams14}.

\subsection{Correlations with the Total Mass-Density Profile Slope}\label{ssec:corrs}
\begin{figure*}
    \includegraphics[resolution=100, width=\textwidth]{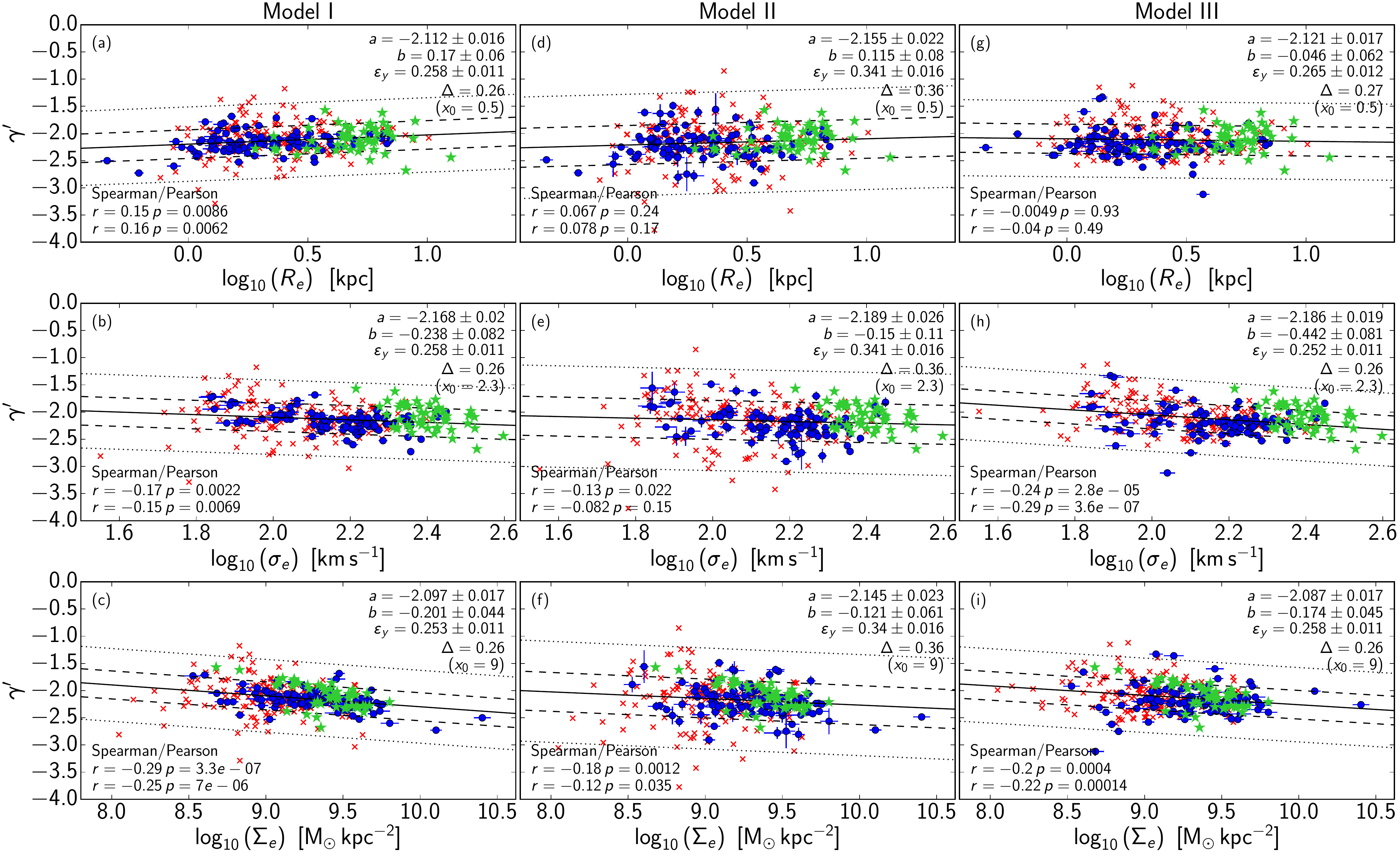}
    \caption{The observed trends of total mass-density profile slope with effective radius (left column), velocity dispersion within an effective radius (middle column), and total-mass surface density (right column). The rows from top to bottom are the results from \mi, \mii, and \miii, respectively. The galaxies with data quality \(\leq 1\) are represented by red crosses, which were flagged for having strong dust, poor kinematic data, or other features that are detrimental to the modelling \citep[see Table 1,][]{cappellari13a}. The blue points are all other objects from the sample used in this work. The green stars are SLACS data, where \(\gamma^{\prime}\) are published by \protect\cite{auger10}, \(\sigma_e\) is published by \protect\cite{treu10}, \(R_e\) is published by \protect\cite{posacki15}, and the values of \(\Sigma_e\) are computed from the MGEs and \(M/L\) values of \protect\cite{posacki15}. All values from both SLACS and \atd are used in the fits and correlation coefficients. The black solid line is the best-fitting line of the general form \(y = a + b\left(x-x_0\right)\), with \(a\), \(b\), and \(x_0\) inset. The dashed and dotted lines are the formal \(1\) and \(2.6\sigma\) bounds, enclosing \(68\%\) and \(99\%\) of the data, respectively. \(\varepsilon_y\) and \(\Delta\) are the intrinsic and observed scatter, respectively. The Spearman and Pearson coefficients (top and bottom values, respectively) are also inset. The figure is produced using the Python implementation of the \textsc{lts\_linefit} procedure of \protect\cite{cappellari13a}.}
    \label{img:sCorr}
\end{figure*}
Presented here are the constraints on the scaling relations of the total mass-density slopes found from our analysis. \autoref{img:sCorr} shows the trends with effective radius, the velocity dispersion within an effective radius, and the total-mass surface density. The surface density is computed by
    \begin{equation}
        \Sigma_{\rm e} = \frac{M_{\rm JAM}}{2 \pi R_e^2}
    \end{equation}
with \(M_{\rm JAM}\) and \(R_e\) tabulated in \cite{cappellari13a}. Overall, we find that the total slopes are quite universal, in agreement with previous results. Moreover some trends do not appear robust and have some dependence on the particular model assumptions. We observe the most significant correlations from \mi. The increased generality of the assumptions in \mii\ may be the cause of the larger scatter in these correlations, discussed further in \autoref{ssec:dist}.\par

\cite{auger10} report a correlation of greater than \(3\sigma\) significance of the total slope with effective radius. This is in agreement with the models presented by \cite{dutton14}, which show correlations of varying degree. \cite{tortora14b} computed spherical, isotropic mass models with various assumed dark matter haloes, using data from the SPIDER \citep{labarbera10} and \atd ETG samples. They found a very clear trend of the total mass-density profile slope with \(R_e\), with smaller galaxies having steeper slopes. Here, we find a weak correlation of the total slope with \(R_e\) only in \mi. Conversely, both Models {\bf II} and {\bf III} from this work, as well as \cite{cappellari15}, show no significant correlations with \(R_e\).\par
All of our models are suggestive of a negative correlation with \(\sigma_e\) for galaxies with \(\log_{10}\left(\sigma_e\right) < 2.1\). We observe a break at \(\log_{10}(\sigma_e) \approx 2.1\), with a weakly-increasing slope for galaxies that have high velocity dispersion. This is interesting, given that it is present in all of our models. We quantify this impression in \autoref{img:sigloess} for \mi\ (though the break is strikingly similar in all of our models), using a generic broken power-law of the form
\begin{equation}
    \gamma^\prime =
    \begin{cases}
        A_0 \left(\frac{x}{x_s}\right)^{\alpha_1};&x<x_s \\
        A_0 \left(\frac{x}{x_s}\right)^{\alpha_2};&x>x_s
    \end{cases}
    \label{eq:bpl}
\end{equation}
where \(A_0\) is a scale factor, \(x_s\) is the break value, and \(\alpha_1\) and \(\alpha_2\) are the slopes of each regime. To determine the significance of this, we again make use of MCMC to optimise the free parameters of \autoref{eq:bpl}.
\begin{figure}
    \includegraphics[resolution=200, width=\columnwidth]{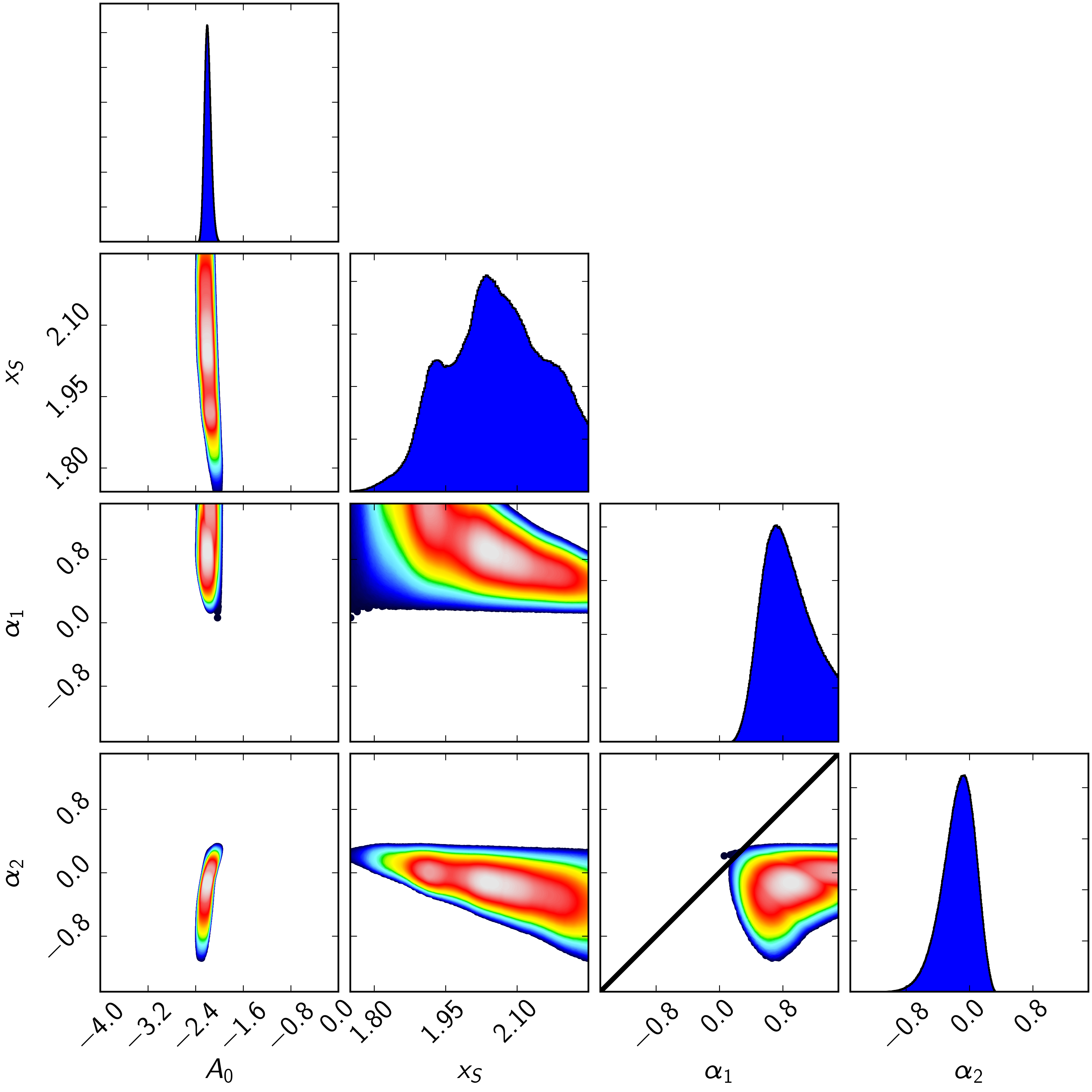}
    \caption{The posterior distributions for the fit of the broken power-law in \autoref{eq:bpl} to \mi. The points are coloured according to their likelihood, with white being the most likely. The extents of each panel illustrates the parameter limits imposed on MCMC. Points falling on the black diagonal line indicate a single power-law (\(\alpha_1 = \alpha_2\)).}
    \label{img:optbpl}
\end{figure}
It is clear from \autoref{img:optbpl} that a single power-law relationship (with \(\alpha_1=\alpha_2\)) is excluded by the posteriors with \(3\sigma\) confidence. Importantly, without any prior bias, the optimal value for \(x_s\) indicates a break at \(\log_{10}(\sigma_e) \approx 2.1\). We explore the significance of this break further in \autoref{ssec:dist}.\par
\begin{figure}
    \includegraphics[resolution=200, width=\columnwidth]{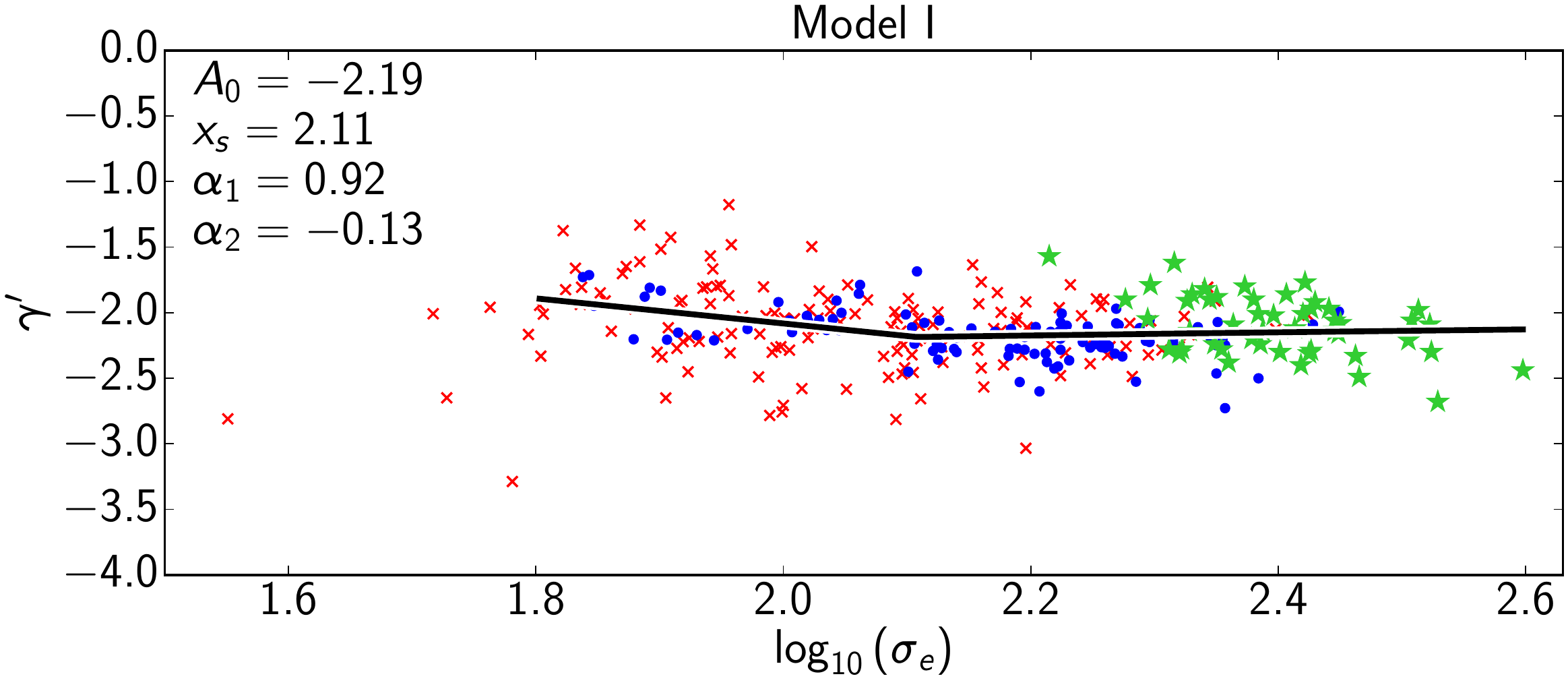}
    \caption{The correlations of the total slope with velocity dispersion (as in \autoref{img:sCorr}h). The blue points are our good data, while the red crosses are for galaxies with data quality \(\leq 1\), which were flagged for having strong dust, poor kinematic data, or other features that are detrimental to the modelling \citep[see table 1 of][]{cappellari13a}. The green stars are SLACS data, where \(\gamma^{\prime}\) is published by \protect\cite{auger10} and \(\sigma_e\) is published by \protect\cite{treu10}. The black curve is the best-fit generic broken power-law given in \autoref{eq:bpl}, which is fit to all SLACS and \atd data, with the optimal values inset.}
    \label{img:sigloess}
\end{figure}
Correlations with \(\sigma_{e/2}\) are also predicted by the models of \cite{dutton14}, which range from weak to strong depending on the model assumptions. Their mass-follows-light model matches best our results when restricted to the region \(\log_{10}(\sigma_e) \ga 2.1\). However, all other models from that work show steeper trends with \(\sigma_{e/2}\), which our data do not support. Moreover, despite the fact that the data presented by \cite{auger10} is at the high end of our \(\sigma_e\) coverage, they find a weak negative correlation, which causes tension with our results. \cite{tortora14b} found tentative evidence for steepening total-mass slopes with increasing effective velocity dispersion, supported by their models that test various dark-matter haloes, across a similar range of \(\sigma_e\) as used here.\par
Finally, the correlation of total slope with total-mass surface density from \mi\ shows a very clear and statistically-significant inverse relationship. All other models in this work, as well as all models from \cite{dutton14} and the data from \cite{auger10}, show similar trends of the total slope with the surface density despite, for instance, the added freedom in \mii\ or the different observational conditions in \cite{auger10}. Furthermore, \cite{sonnenfeld13} studied the evolution of structure in ETGs over a modest redshift range using lensing observations. They model their objects with a spherical single power-law as done by \cite{auger10}. Their results show definite correlations between the total mass-density profile slope and the \emph{stellar} surface density, with a rate of \(\partial \gamma^\prime\big/\partial \Sigma_{\star} = -0.38 \pm 0.07\) (accounting for their reversed definition of \(\gamma^\prime\)). This is steeper than the trends shown in \autoref{img:sCorr} (the \(b\) parameter), though this is likely due to the differences between total and stellar surface densities within \(1~R_e\). \cite{newman15} conducted a consistent analysis of ETG lenses, ranging from galaxy-scale objects (taken from SLACS) to cluster-scale lenses. Their results agree with those of \cite{sonnenfeld13}, reporting a value of \(\partial \gamma^\prime\big/\partial \Sigma_{\star} = -0.34 \pm 0.11\). They also find that the total slope depends significantly on the halo mass, \(M_{200}\), reporting \(\partial \gamma^\prime\big/\partial M_{200} = 0.33 \pm 0.07\). It is perhaps not surprising, then, in light of these dependencies and the size of our galaxy sample, that we find a larger scatter in the total slopes compared to previous work.\par
\begin{figure}
    \includegraphics[resolution=600, width=1\columnwidth]{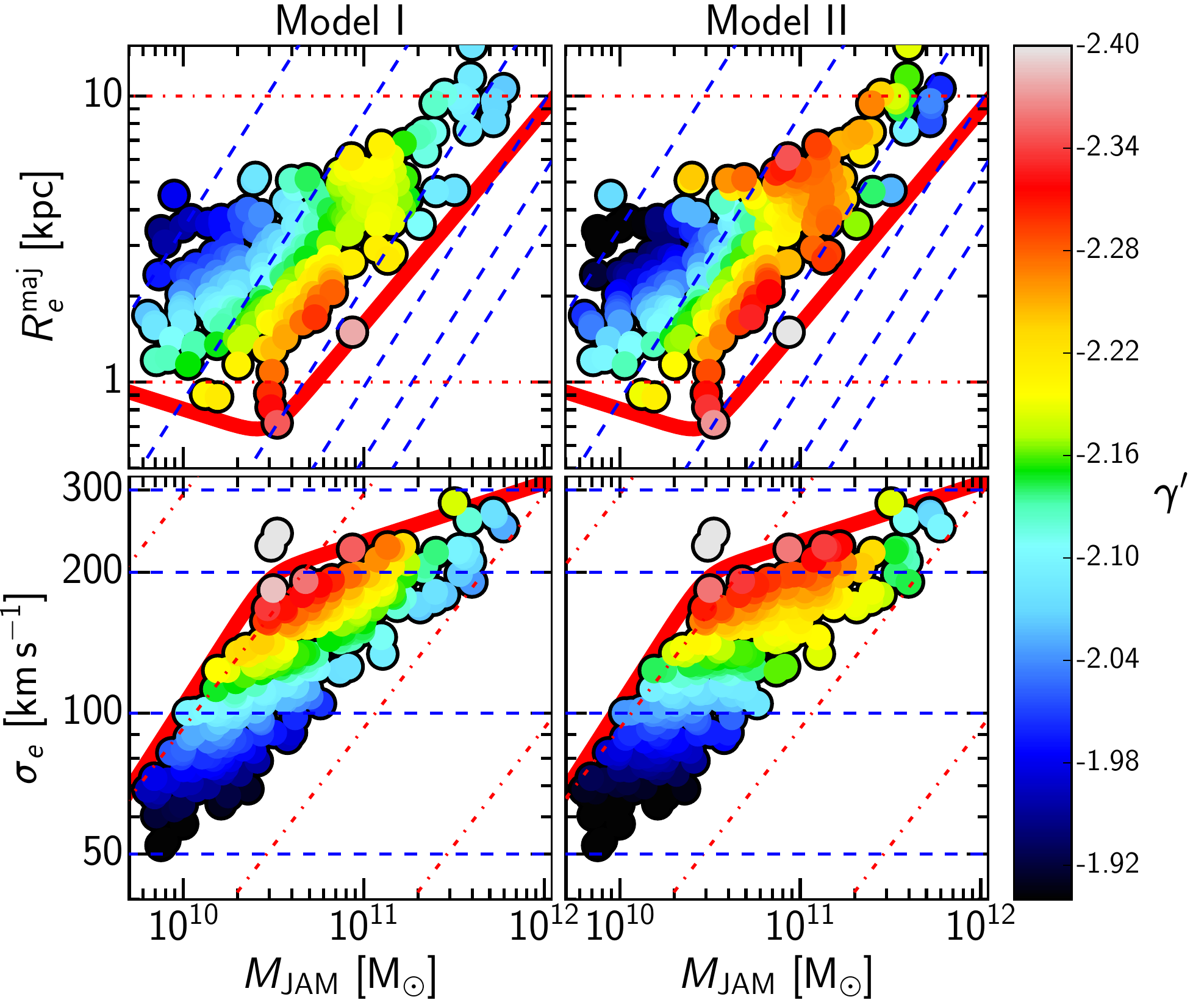}
    \caption{The correlations between total slope, dynamical mass, effective radius, and velocity dispersion within an effective radius. The total slope measurements determined by \mi\ are in the left column, while those from \mii\ are in the right column. The corresponding figure for \miii\ was published as Fig. 22c in \protect\cite{cappellari16}. The mass, radius, and velocity dispersion are those published by \protect\cite{cappellari13a}. The measure of \(R_e\) used in this relation is the length of the major axis of the isophote containing half of the light of the MGE model. The points are coloured according to mean logarithmic \emph{total}-mass slope, which have been smoothed by the \textsc{loess} algorithm. The blue dashed lines are lines of constant velocity dispersion, where \(\sigma_e = [50, 100, 200, 300, 400, 500]~\si{km s^{-1}}\). The dot-dashed red lines are lines of constant effective radius, where \(R_e^{\rm maj} = [0.1, 1, 10, 100]~\si{kpc}\). The thick red line is the ZOE of \protect\cite{cappellari13b}.}
    \label{img:msp}
\end{figure}

In addition to the galactic observables tested above, we investigate the total slope on the mass plane \citep{cappellari13a}, presented in \autoref{img:msp}.
The thick red line is the Zone of Exclusion (ZOE), given by Eq. 4 of \cite{cappellari13b}. The coloured points have been smoothed by the Locally Weighted Regression technique (\textsc{loess}) of \cite{cleveland79} as implemented in the \texttt{loess\_2d}\footref{fn:atlas} Python procedure of \cite{cappellari13a}. The aim of \textsc{loess} is to characterise the underlying distribution of the data, even with relatively small samples. We observe that, below the critical stellar mass \(M_{\mathrm{crit}} \approx 2\times 10^{11}~\si{\Msun}\), the total slope follows lines of constant velocity dispersion in all panels, while the slope decreases above \(M_{\mathrm{crit}}\), which is consistent with Fig. 22c of \cite{cappellari16} using \miii. This agrees with the correlations detected in \autoref{img:sCorr}.

\subsection{Distributions of Mass-Density Profile Slopes}\label{ssec:dist}

\begin{figure}
    \includegraphics[resolution=600, width=1\columnwidth]{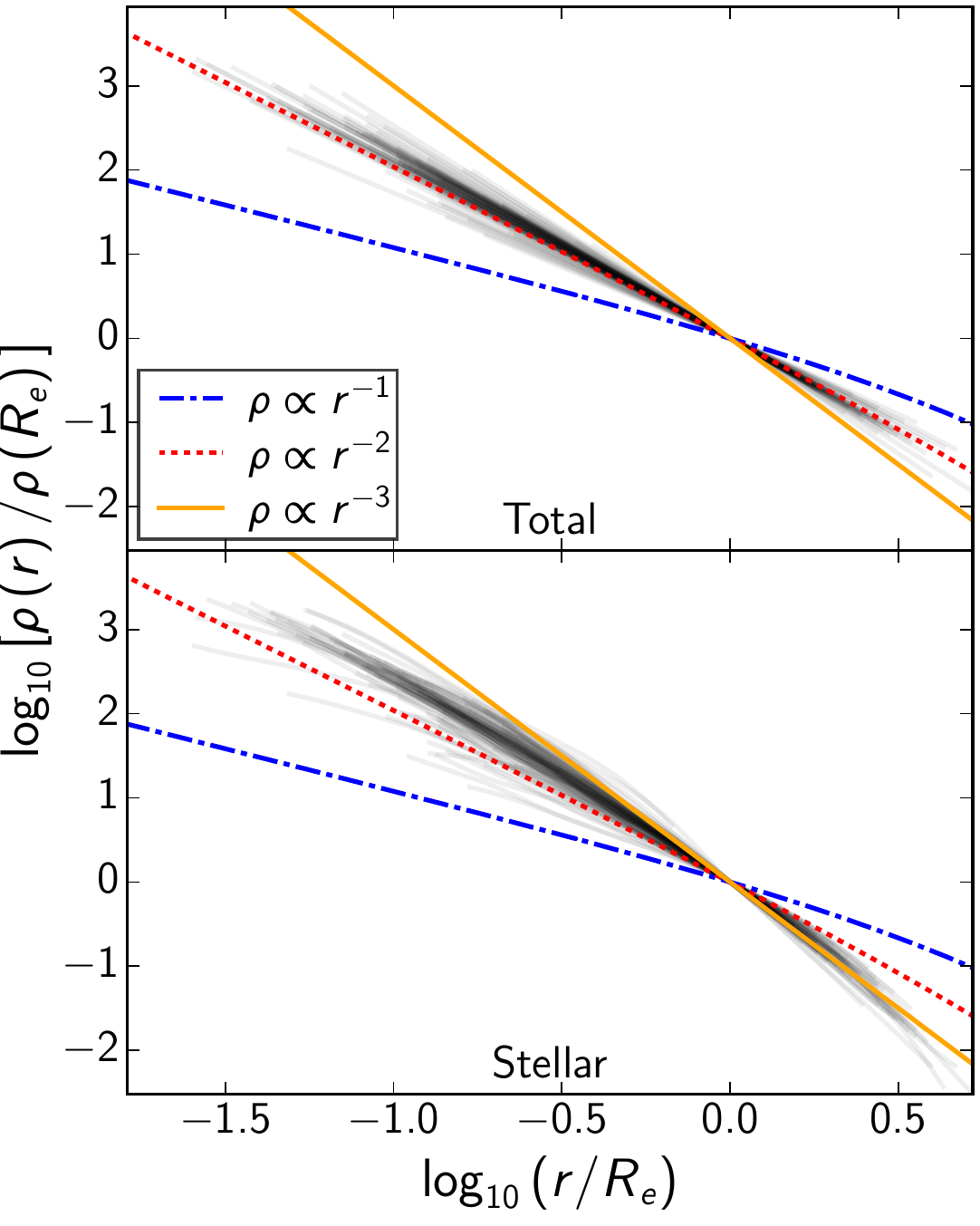}
    \caption{The profiles (as computed in \mi) of the total (top panel) and stellar (bottom panel) mass density. Very high transparency is used here to emphasise the highest density of curves. The total mass-density profiles remain nearly isothermal across the entire radial range. The stellar profiles are mildly steeper than isothermal in the inner region, then steepen further in the outer regions. All stellar profiles have been scaled by their corresponding \ml{Salp} profiles. Each curve is plotted on the fitted interval \([2\si{\arcsecond}, r_{\rm max}]\), so that any visible trends are based on constrained data, rather than extrapolated curves. This is also why the curves have different lengths.}
    \label{img:slopes}
\end{figure}

The total and stellar profiles for the full \atd sample are presented in \autoref{img:slopes}.
It can be seen that the total slopes remain nearly-isothermal across the entire radial range of the data, and that indeed the spread of slopes is small. The stellar slopes in this figure are those found by converting the luminosity into mass using the \ml{Salp}\((R)\) profile, namely assuming a Salpeter IMF. The stellar density profiles steepen in the outer regions, such that \(\rho_{\mathrm{star}}(r>R_e)\propto r^{-3}\). These results are in very good agreement with \cite{cappellari15}, who find similar behaviour in both the total and stellar density profiles within similar radial ranges.\par
\begin{figure}
    \centerline{
    \includegraphics[resolution=600, width=\columnwidth]{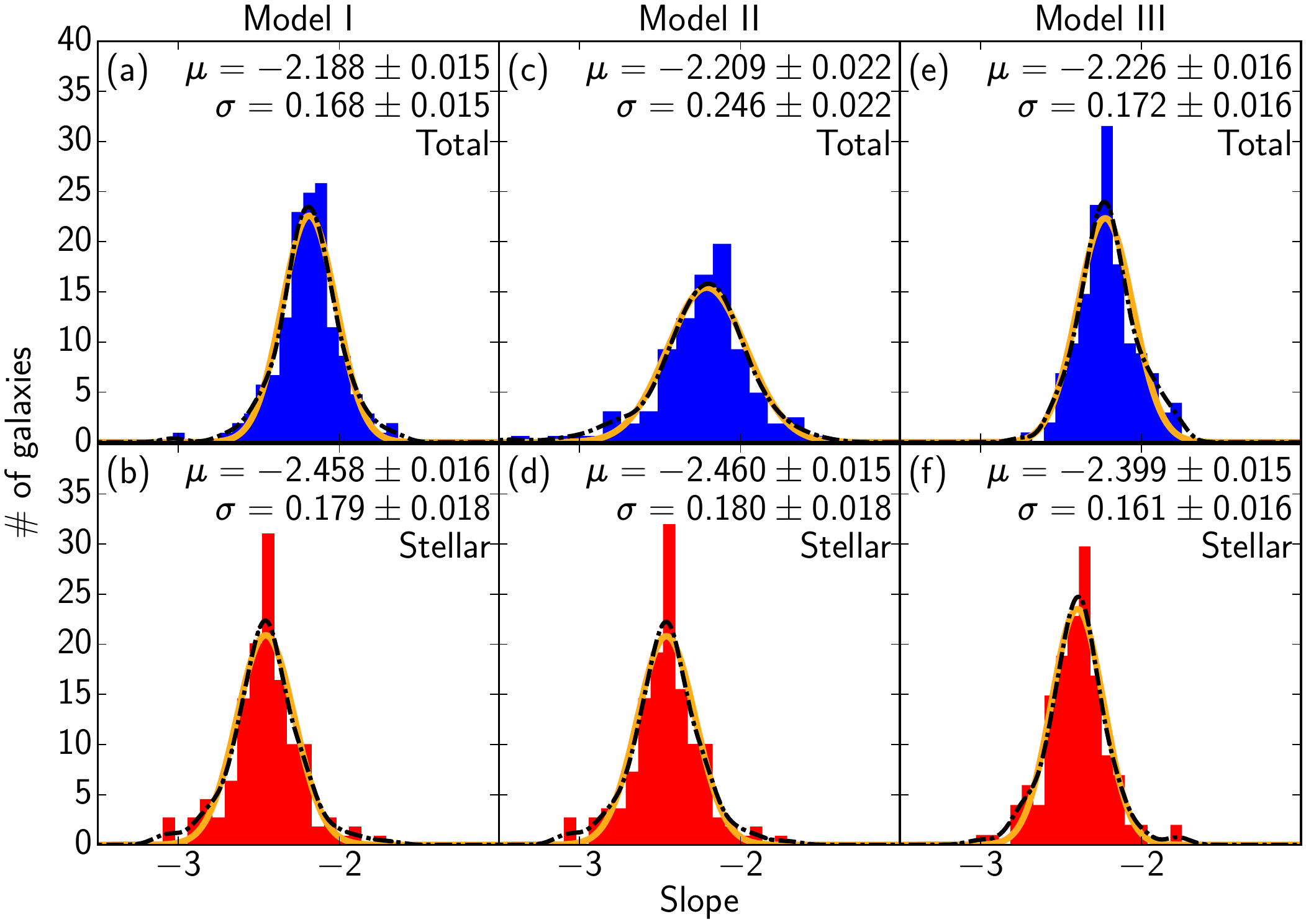}}
    \caption{The distribution of the mass-density profile slopes for the 142 galaxies from the \atd sample with \(\log_{10}(\sigma_e) > 2.1\). The black dot-dashed line is a kernel density estimation of the underlying distribution, using a Gaussian kernel. The orange solid line is a Gaussian fit to this kernel density estimation. The parameters and errors of the fitted Gaussians are inset. These values can be compared directly to the biweight estimates in \autoref{tab:slopes}. The left, centre, and right columns are Models \textbf{I}, \textbf{II}, and \textbf{III}, respectively. The top and bottom rows are the mass-density profile slopes of the total and stellar-only components, respectively.}
    \label{img:allSlopes}
\end{figure}

Based on our findings in \autoref{ssec:corrs}, we investigate the distribution of slopes for galaxies with \(\log_{10}(\sigma_e) > 2.1\). This is to avoid any variations of the total slopes with other galactic observables that would skew or otherwise alter the one-dimensional histogram. These distributions are presented in \autoref{img:allSlopes}.\par
We observe the largest scatter in the distribution of total slopes from \mii. To understand why, we studied in detail some galaxies where the total slope was significantly different between \mii\ and the other two models. We found that in those cases the large differences were due to the fact that \mii\ tried to use the extra freedom in the total mass-density profile parametrization to fit the complex kinematics of barred galaxies. This suggests that \mii\ is less robust than the others against systematic issues in the modelling assumptions. For this reason we ignore the results of \mii\ in our main conclusions.\par
The scatter reported in this work is computed as the biweight sigma, which is shown to be more robust to outliers and small-number statistics compared to, for instance, the conventional standard deviation \citep{beers90}. Errors on the mean and variance estimators are computed using bootstrapping techniques. We repeatedly draw a random subset of total slopes from the sample, and compute the mean and biweight sigma. We then find the variance of these subsets of means and biweight sigmas. In addition, as seen in \autoref{img:allSlopes}, we compute kernel density estimations for all distributions, assuming a Gaussian kernel. We then fit a Gaussian to this density estimation in order to gauge how well the distributions are described by Gaussians, and consequently which variance estimators are appropriate.\par
Presented in \autoref{tab:slopes} are the properties of our distributions, along with the results of other works.
\begin{table}
    \caption{The total mass-density profile slopes.}
    \label{tab:slopes}
\begin{threeparttable}
    \begin{center}
    \(\begin{tabu}{cccc}
        \toprule
        \gamma^\prime_{\rm tot} & \sigma_{\gamma^\prime}^\mathrm{tot} & {\rm median} & \text{Reference}\\
        && r/R_e &\\\midrule
        \slTVal \pm \slTValErr & \biTVal \pm \biTValErr & 0.9 & \text{\mi, SD}\\
        \slTValSev \pm \slTValErrSev & \biTValSev \pm \biTValErrSev & 0.9 & \text{\mii, SD}\\
        \slTValAtl \pm \slTValErrAtl & \biTValAtl \pm \biTValErrAtl & 0.9 & \text{\miii, SD}\\
        -2.078 \pm 0.027 & 0.22 & 0.5 & \text{\protect\cite{auger10}, L}\\
        -2.19 \pm 0.03 & 0.11 & 4 & \text{\protect\cite{cappellari15}, SD}\\
        -2.18 \pm 0.03 & 0.11 & 6 & \text{\protect\cite{serra16}, GD}\\
        \bottomrule
    \end{tabu}\)
    \end{center}
    \begin{tablenotes}[para, flushleft]
        \note The results here are for the 142 galaxies from the \atd sample with \(\log_{10}(\sigma_e) > 2.1\). The (observed) scatter reported for our models is given by the biweight sigma of \protect\cite{beers90}. In addition, we compute the observed scatter using biweight sigma from the data of \protect\cite{auger10}. The \(r\) column contains the median radius probed by each work, in units of \(R_e\). The type of observation used is provided with the reference, where SD is stellar dynamics, L is lensing, and GD is gas dynamics.
    \end{tablenotes}
\end{threeparttable}
\end{table} Again, these results show close agreement with previous work, which use different observational techniques and samples. There is a clear tendency towards an isothermal total slope for the ETGs studied in the above works. By virtue of the models run in this work, our results are free from any assumptions or biases about the stellar population or dark matter (these are introduced at a later stage of the analysis), since we model the total mass directly. This is similarly the case with the data from \cite{auger10}, though with less-detailed models.\par
We similarly characterise the slopes of the stellar-only component within each galaxy. These results are presented in \autoref{tab:sslopes}.
\begin{table}
\caption{The stellar-only mass-density profile slopes.}
    \label{tab:sslopes}
\begin{threeparttable}
    \begin{center}
    \(\begin{tabu}{ccc}
        \toprule
        \gamma^\prime_{\rm star} & \sigma_{\gamma^\prime}^\mathrm{star} & \text{Model}\\\midrule
        \slSVal \pm \slSValErr & \biSVal \pm \biSValErr & \text{radial \ml{Salp}}\\
        -2.400 \pm 0.016 & 0.169 \pm 0.015 & \text{constant }M/L\\
        \bottomrule
    \end{tabu}\)
    \end{center}
    \begin{tablenotes}[para, flushleft]
        \note Radial \ml{Salp} models have the stellar luminosity converted to mass using the derived \ml{Salp}\((R)\) profiles. The stellar slopes of the constant-\((M/L)\) model were computed from the MGE fits to the surface brightness of \protect\cite{scott13} deprojected at the inclination determined by \protect\cite{cappellari13a}. This corresponds to \miii.
    \end{tablenotes}
\end{threeparttable}
\end{table} The inclusion of the \ml{Salp} profiles during the decomposition of the stellar and dark mass components has only a small impact on the resulting distribution of stellar slopes. While the peak is shifted to steeper slopes when taking into account the stellar populations, the magnitude of the shift is within the observed scatter.\par
\begin{figure}
    \centerline{
    \includegraphics[resolution=600, width=\columnwidth]{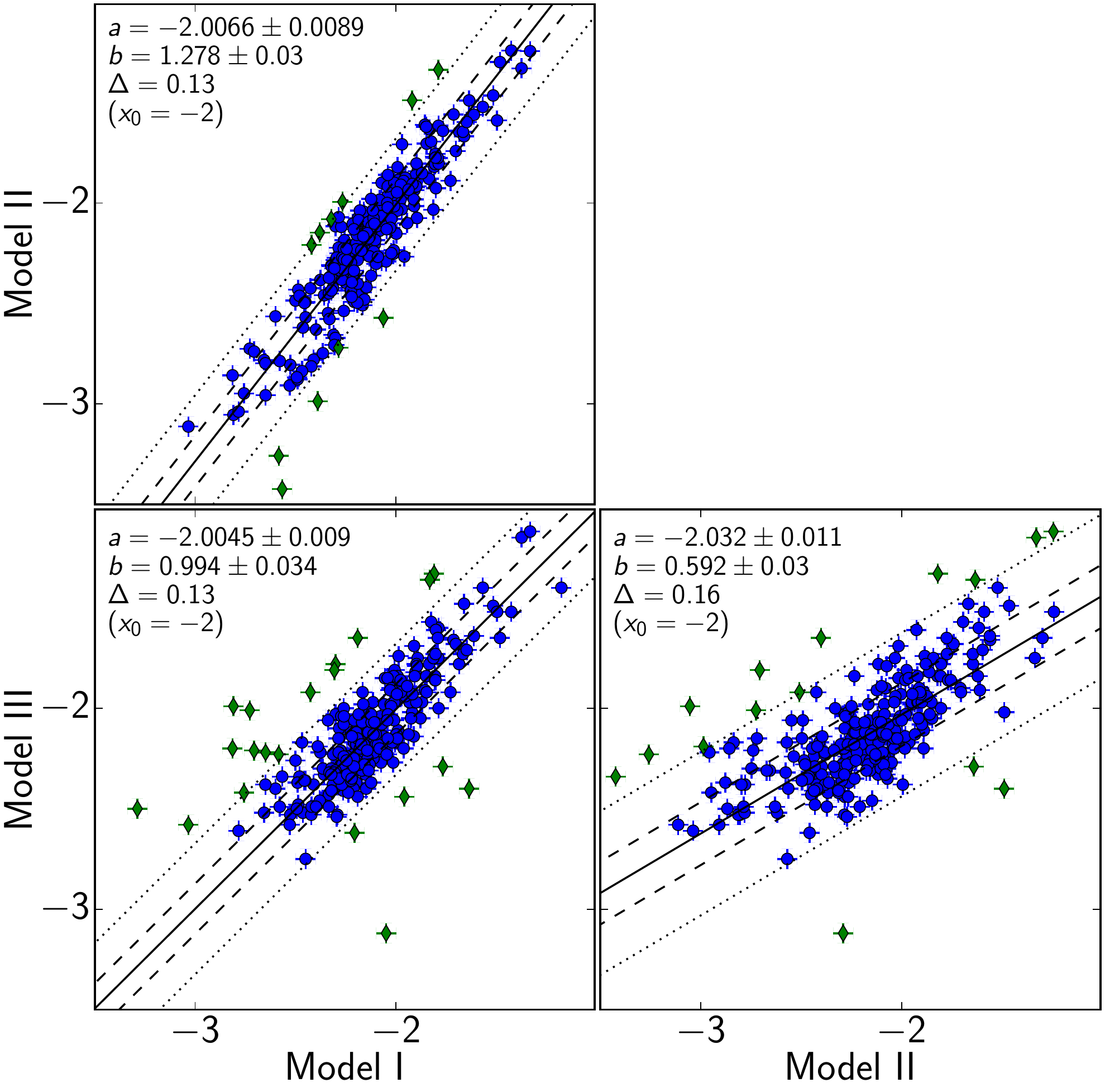}}
    \caption{Comparisons between every model in this work of the total-mass density profile slope. The black solid line is the best-fitting line of the general form \(y = a + b\left(x-x_0\right)\), with \(a\), \(b\), and \(x_0\) inset. The dashed and dotted lines are the formal \(1\) and \(2.6\sigma\) bounds, enclosing \(68\%\) and \(99\%\) of the data, respectively. \(\Delta\) is the observed scatter. The figure is produced using the Python implementation of the \textsc{lts\_linefit} procedure of \protect\cite{cappellari13a}. The green diamonds are galaxies that are considered outliers by \textsc{lts\_linefit}, and the blue points are all other objects.}
    \label{img:battle}
\end{figure}

Our posterior distributions indicate small errors in our total slopes, but the true errors are likely dominated by systematic uncertainties in the modelling assumptions, rather than by purely statistical errors. The availability of these different sets of models allows us to perform a more conservative estimate of the errors, including these systematics, by comparing the results between the three different models. This comparison is shown in \autoref{img:battle}.
The data sets from all models are compared using \textsc{lts\_linefit}, which subsequently estimates the intrinsic scatter. We find that \mi\ agrees with both \mii\ and \miii, with an observed scatter of \(\Delta=0.13\). As we are measuring the same quantity on both axes, we attribute this scatter to our measurements uncertainties, including systematic errors. Assuming these are comparable on both axes, we infer a measurement error of \(\varepsilon_{\gamma^\prime}=\Delta/\sqrt{2}=0.09\) on our total slopes. We additionally find that the correlation between \mi\ and \miii\ has a slope of unity, within the errors, while both correlations involving \mii\ are significantly different from unity. This provides further evidence for the negative impact of the systematic issues on \mii.\par
Both the mean and scatter of this distribution agree quite well with the determinations within \(R_e/2\) by \cite{auger10}. They are also in remarkable agreement with the independent measurements at the much larger radii of \(4\) and \(6~R_e\) from \cite{cappellari15} and \cite{serra16}, respectively. We see that the scatter in both the total and stellar distributions remains relatively small across all models, given our sample size. Adopting the result of either \mi\ or \miii, from the observed scatter of \(\sigma_{\gamma^\prime}^\mathrm{tot} \approx \biTVal\) and the measured errors of \(\varepsilon_\gamma^\prime\approx 0.09\), we infer an intrinsic scatter of \(\sigma_{\gamma^\prime}^\mathrm{intr} \approx 0.15\). This value is smaller, but consistent within the errors with the smallest scatter \(\sigma_{\gamma^\prime}^\mathrm{star} \approx 0.169\) we measure directly from the photometry for the constant-\(M/L\) model. The scatter is significantly smaller than \(\sigma_{\gamma^\prime}^\mathrm{star} \approx \biSVal\) that we measure for the stellar profiles when accounting for \(M/L\) variations caused by age and metallicity gradients.\par
These results indicate the following:
\begin{enumerate*}[label=(\roman*)]
    \item remarkably, the accuracy in our total slopes, using complex dynamical models, is at least comparable to the accuracy in measuring stellar slopes directly from the photometry
    \item the addition of dark matter to the stellar densities does not increase the scatter in the density slopes. This is not inconsistent with the previously claimed conspiracy between the stellar and dark components. However, we cannot make strong claims in this direction as the dark matter contributes only a small fraction of the total mass within the region we sample with our kinematics. For this reason, even a perfect conspiracy would have a minor effect on our scatter.
\end{enumerate*}

\section{Conclusions}\label{sec:concl}

We have investigated the dark matter content and total mass-density profiles of ETGs covering a broad range of stellar masses in the local Universe. We have compared model assumptions concerning the spatial arrangement of matter and the stellar populations within early-type galaxies. We found generally consistent results for the dark matter fractions using different assumptions.\par
We have characterised the distribution of total and stellar mass-density profile slopes. We found that the slopes of the total-mass profiles tend towards a nearly-isothermal value of \(\gamma^\prime_\mathrm{tot} = \slTVal \pm \slTValErr\) with an observed scatter of only \(\sigma_{\gamma^\prime}^\mathrm{tot}=\biTVal \pm \biTValErr\), which is even smaller than previous work with strong lensing at similar radii. We found the scatter of the total mass-density profile slopes to be marginally smaller than what we measured for the stellar-only profiles. This illustrates that the accuracy of our measurements of the total mass-profile slopes from the kinematics is at least comparable to the measurements of the stellar slopes from the photometry. The inclusion of dark matter does not increase the scatter of the total mass-density profile slopes, which is consistent with the previous claims of a bulge-halo conspiracy.\par
We have studied possible correlations of the total-mass slopes with galactic observables. We found that the most significant correlations of the total slope are those with mass surface density, \(\Sigma_e\), and effective velocity dispersion, \(\sigma_e\). Interestingly, we found evidence of a break in the relation between the total slope and the velocity dispersion. The total slope varies with velocity dispersion for \(\log_{10}(\sigma_e) \la 2.1\) and is nearly universal for \(\log_{10}(\sigma_e) \ga 2.1\). Similarly, we confirmed that the total slopes for masses below \(M_{\mathrm{crit}} \la 2\times 10^{11}~\si{\Msun}\) follow lines of constant velocity dispersion, as illustrated on the mass plane and found by previous authors.\par
This work has shown that the effects of the stellar population and assumptions about the dark matter do not dramatically alter the distributions of the mass-density profile slopes. We have shown that the implementation of our total-mass methodology is able to accurately reproduce observed IFU kinematics, making it a generally-applicable method well suited to the newest generation of large IFU surveys, such as SAMI \citep{croom12} and MaNGA \citep{bundy15}.

\section*{Acknowledgements}

AP thanks Macquarie University for support and financial assistance during this project. MC acknowledges support from a Royal Society University Research Fellowship. RMcD is the recipient of an Australian Research Council Future Fellowship (project number FT150100333). The authors thank the \atd team for making the data publicly available, and the referee for the helpful comments and suggestions.




\bibliographystyle{mnras}
\bibliography{mres2}


%
\clearpage
\onecolumn
\appendix

\section{Parameters}

\input{append.tex}
%


\bsp	
\label{lastpage}
\end{document}

%% file: thumbs.tex
\centerline{
	\includegraphics[resolution=72, height=0.19\textheight]{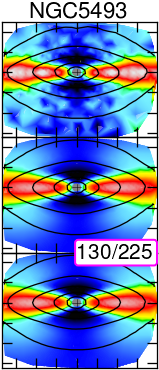}
	\includegraphics[resolution=72, height=0.19\textheight]{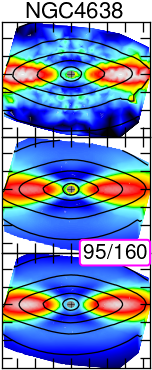}
	\includegraphics[resolution=72, height=0.19\textheight]{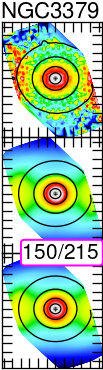}
	\includegraphics[resolution=72, height=0.19\textheight]{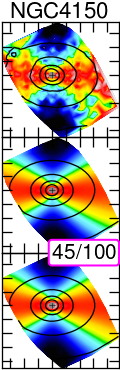}
	\includegraphics[resolution=72, height=0.19\textheight]{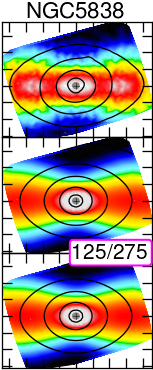}
	\includegraphics[resolution=72, height=0.19\textheight]{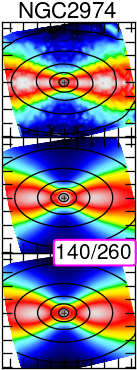}
	\includegraphics[resolution=72, height=0.19\textheight]{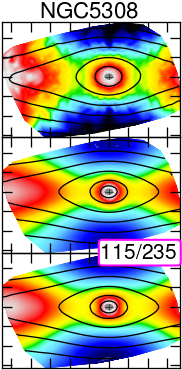}
	\includegraphics[resolution=72, height=0.19\textheight]{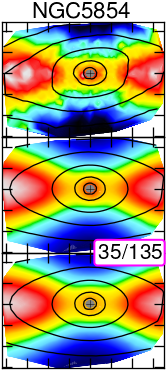}
	\includegraphics[resolution=72, height=0.19\textheight]{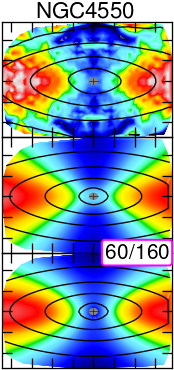}
}
\centerline{
	\includegraphics[resolution=72, height=0.19\textheight]{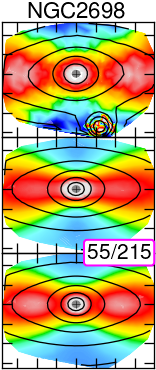}
	\includegraphics[resolution=72, height=0.19\textheight]{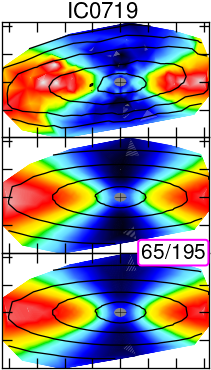}
	\includegraphics[resolution=72, height=0.19\textheight]{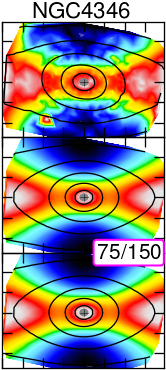}
	\includegraphics[resolution=72, height=0.19\textheight]{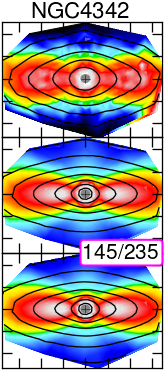}
	\includegraphics[resolution=72, height=0.19\textheight]{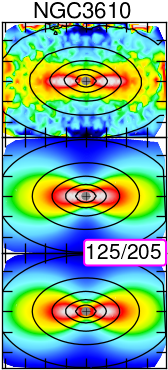}
	\includegraphics[resolution=72, height=0.19\textheight]{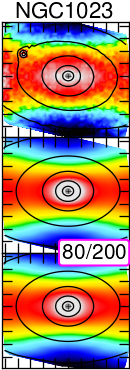}
	\includegraphics[resolution=72, height=0.19\textheight]{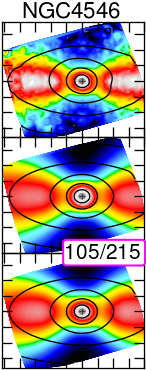}
	\includegraphics[resolution=72, height=0.19\textheight]{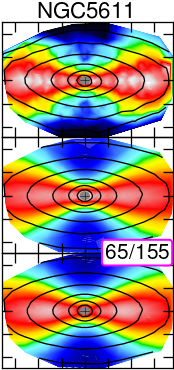}
}
\centerline{
	\includegraphics[resolution=72, height=0.19\textheight]{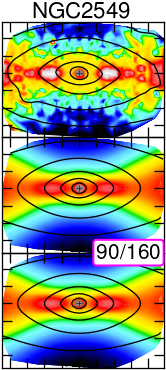}
	\includegraphics[resolution=72, height=0.19\textheight]{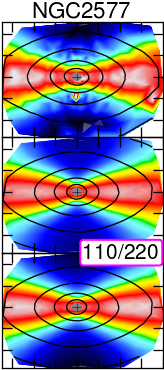}
	\includegraphics[resolution=72, height=0.19\textheight]{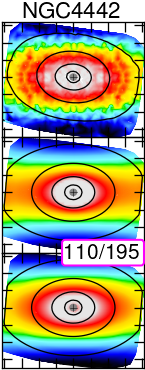}
	\includegraphics[resolution=72, height=0.19\textheight]{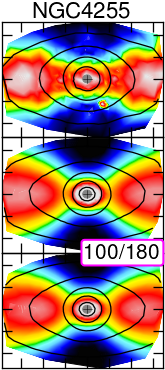}
	\includegraphics[resolution=72, height=0.19\textheight]{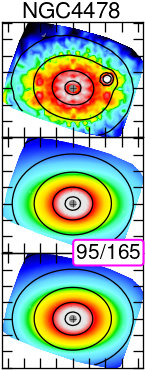}
	\includegraphics[resolution=72, height=0.19\textheight]{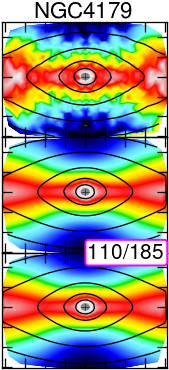}
	\includegraphics[resolution=72, height=0.19\textheight]{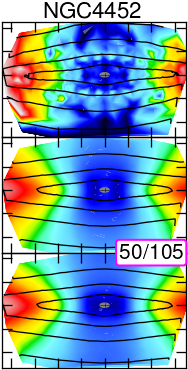}
}
\centerline{
	\includegraphics[resolution=72, height=0.19\textheight]{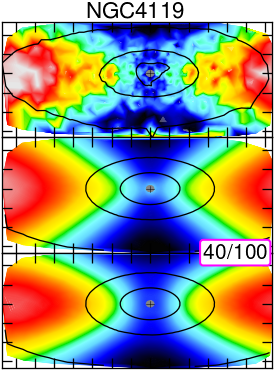}
	\includegraphics[resolution=72, height=0.19\textheight]{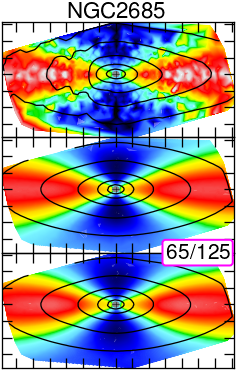}
	\includegraphics[resolution=72, height=0.19\textheight]{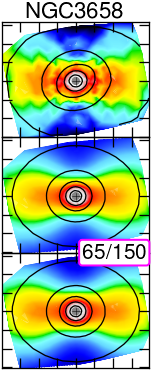}
	\includegraphics[resolution=72, height=0.19\textheight]{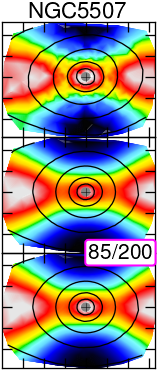}
	\includegraphics[resolution=72, height=0.19\textheight]{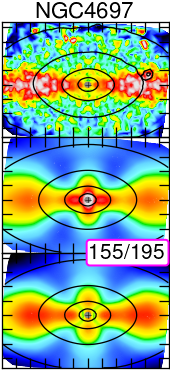}
	\includegraphics[resolution=72, height=0.19\textheight]{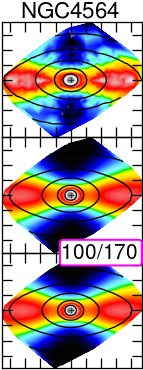}
	\includegraphics[resolution=72, height=0.19\textheight]{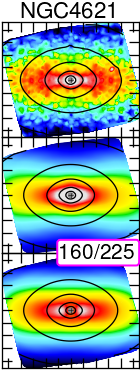}
}
\centerline{
	\includegraphics[resolution=72, height=0.19\textheight]{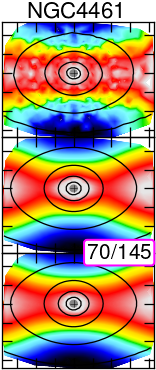}
	\includegraphics[resolution=72, height=0.19\textheight]{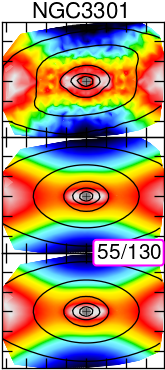}
	\includegraphics[resolution=72, height=0.19\textheight]{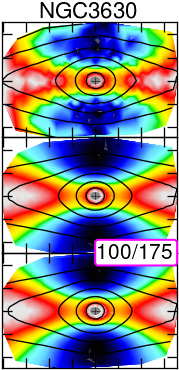}
	\includegraphics[resolution=72, height=0.19\textheight]{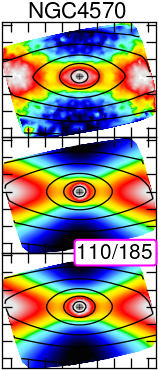}
	\includegraphics[resolution=72, height=0.19\textheight]{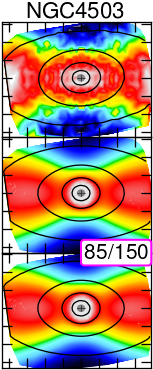}
	\includegraphics[resolution=72, height=0.19\textheight]{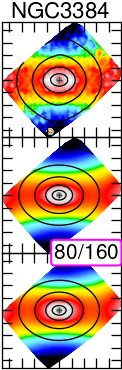}
	\includegraphics[resolution=72, height=0.19\textheight]{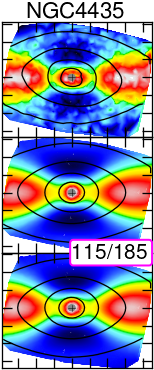}
	\includegraphics[resolution=72, height=0.19\textheight]{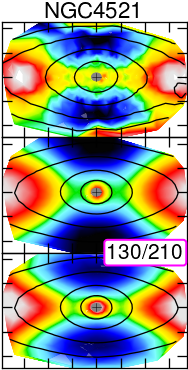}
}

%% file: corn-thumbs.tex
\centerline{
	\includegraphics[resolution=150, height=0.311\textheight]{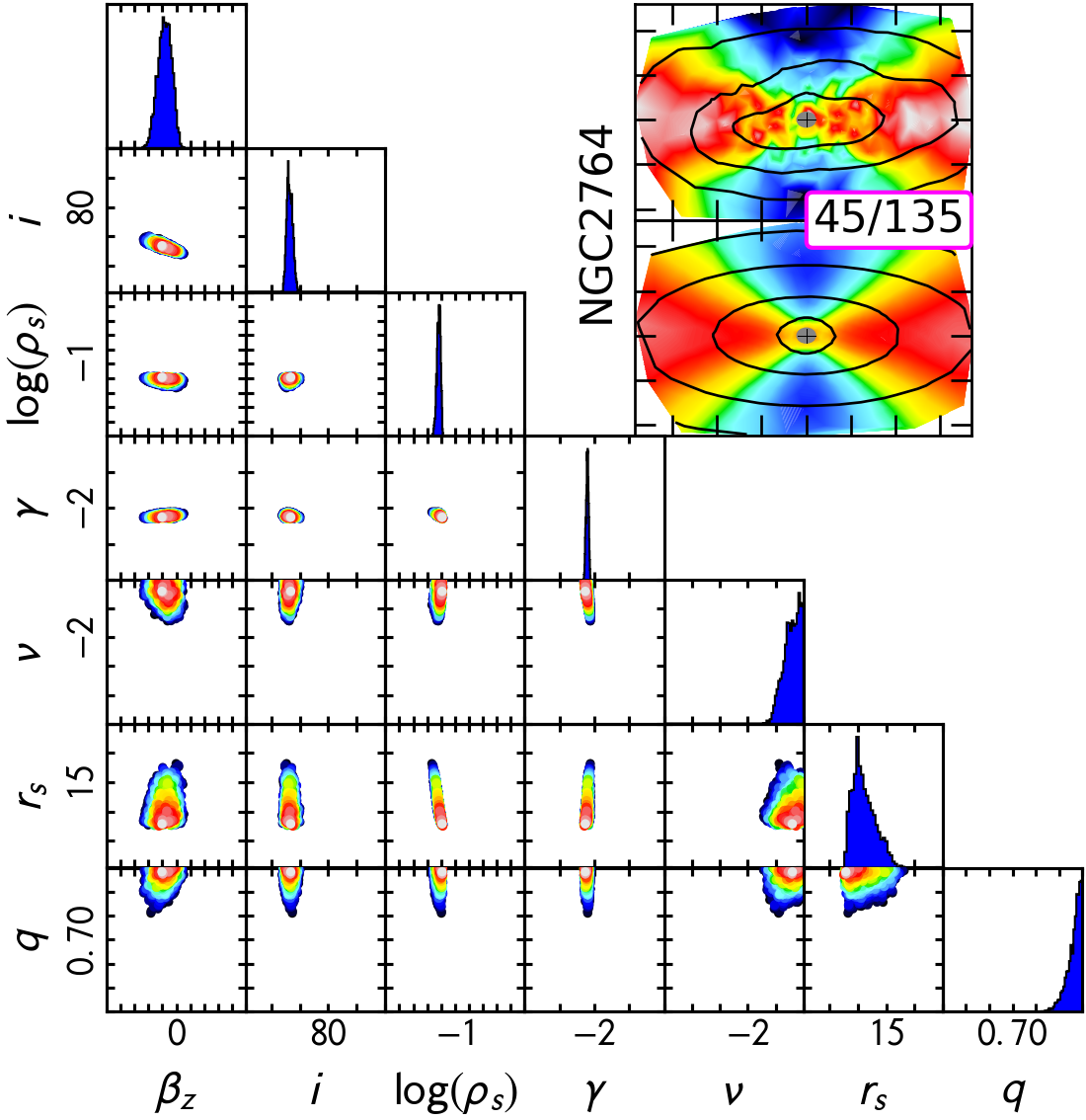}
	\hspace{5em}
	\includegraphics[resolution=150, height=0.311\textheight]{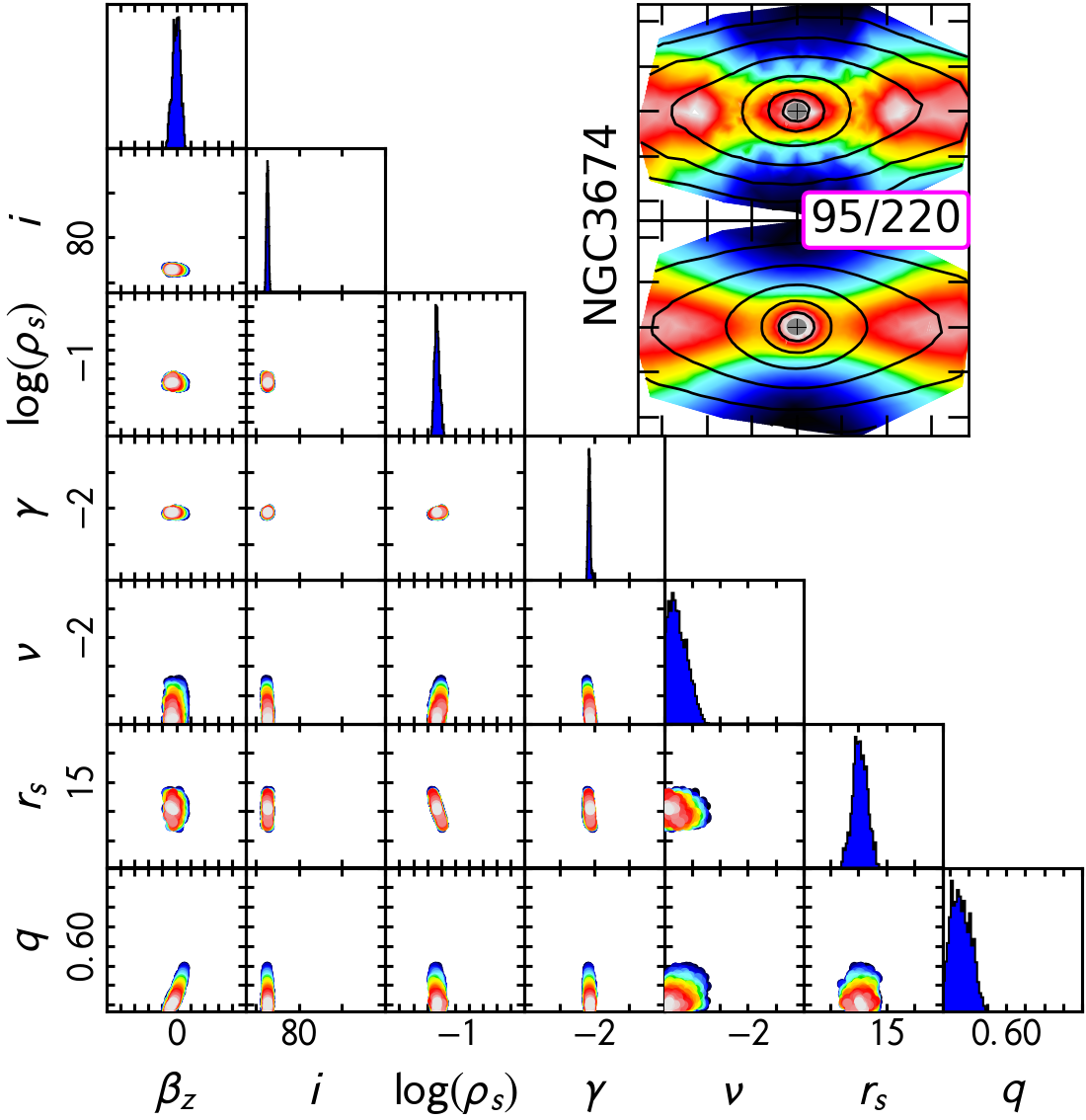}
}\centerline{
	\includegraphics[resolution=150, height=0.311\textheight]{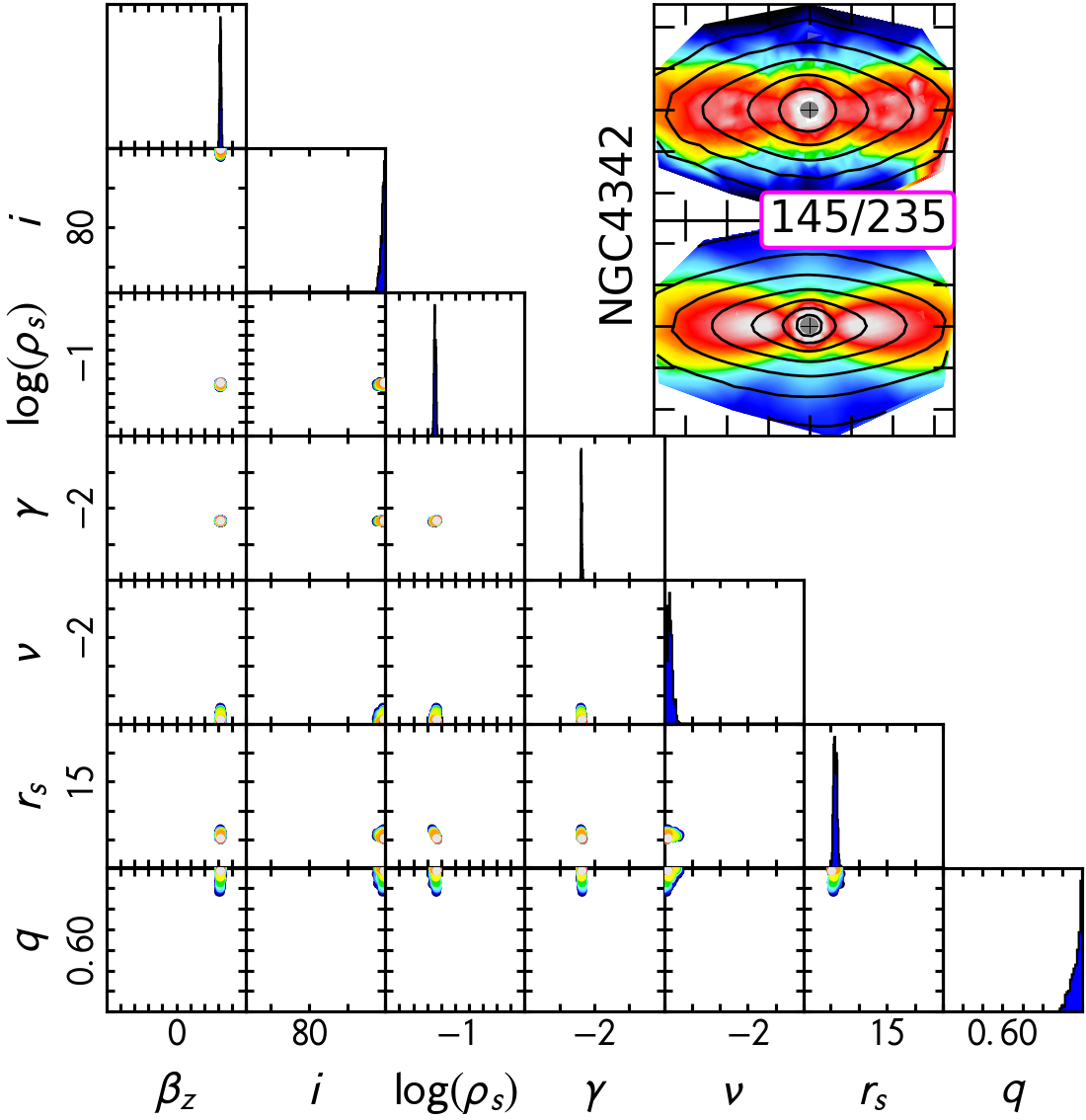}
	\hspace{5em}
	\includegraphics[resolution=150, height=0.311\textheight]{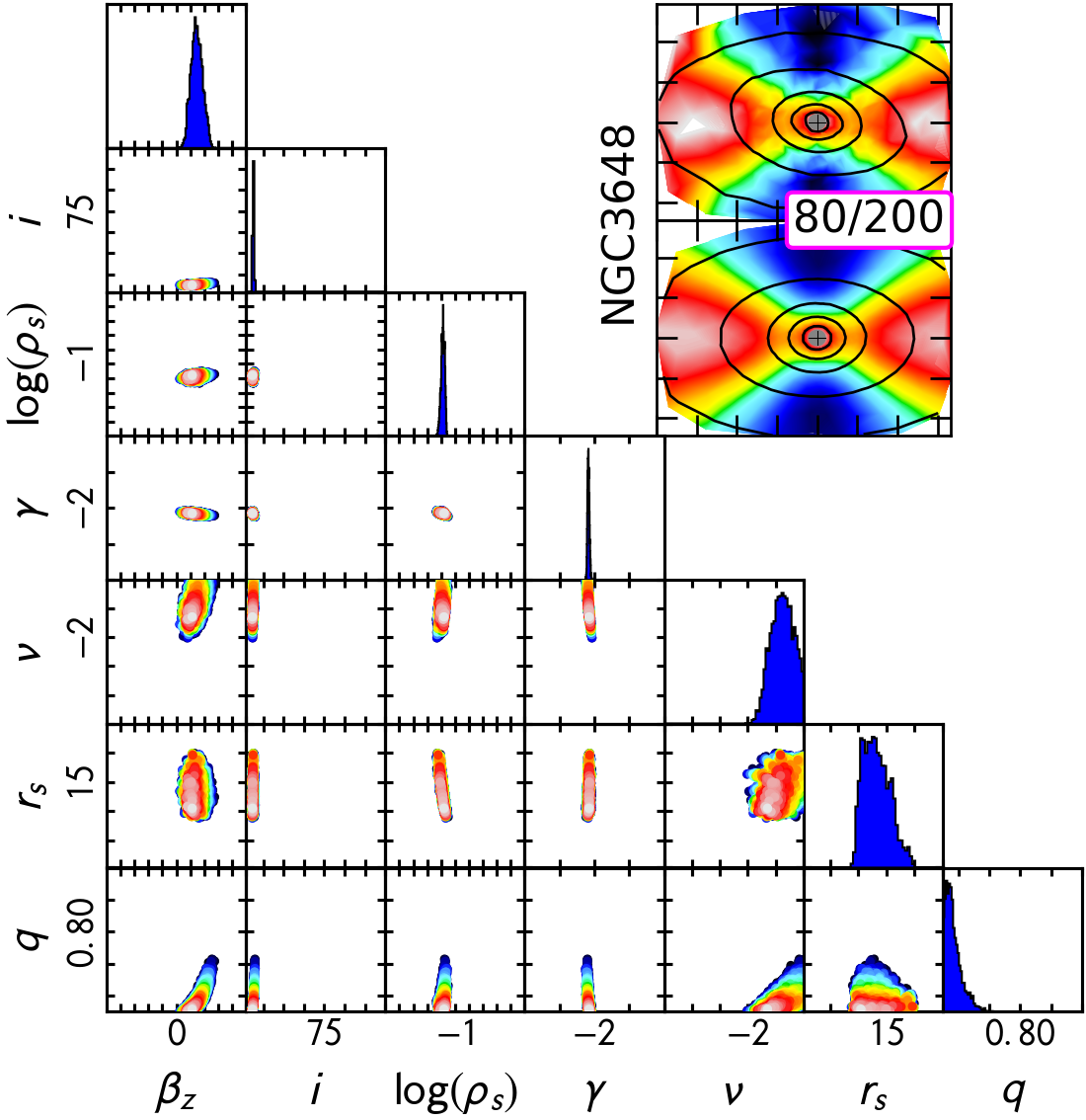}
}\centerline{
	\includegraphics[resolution=150, height=0.311\textheight]{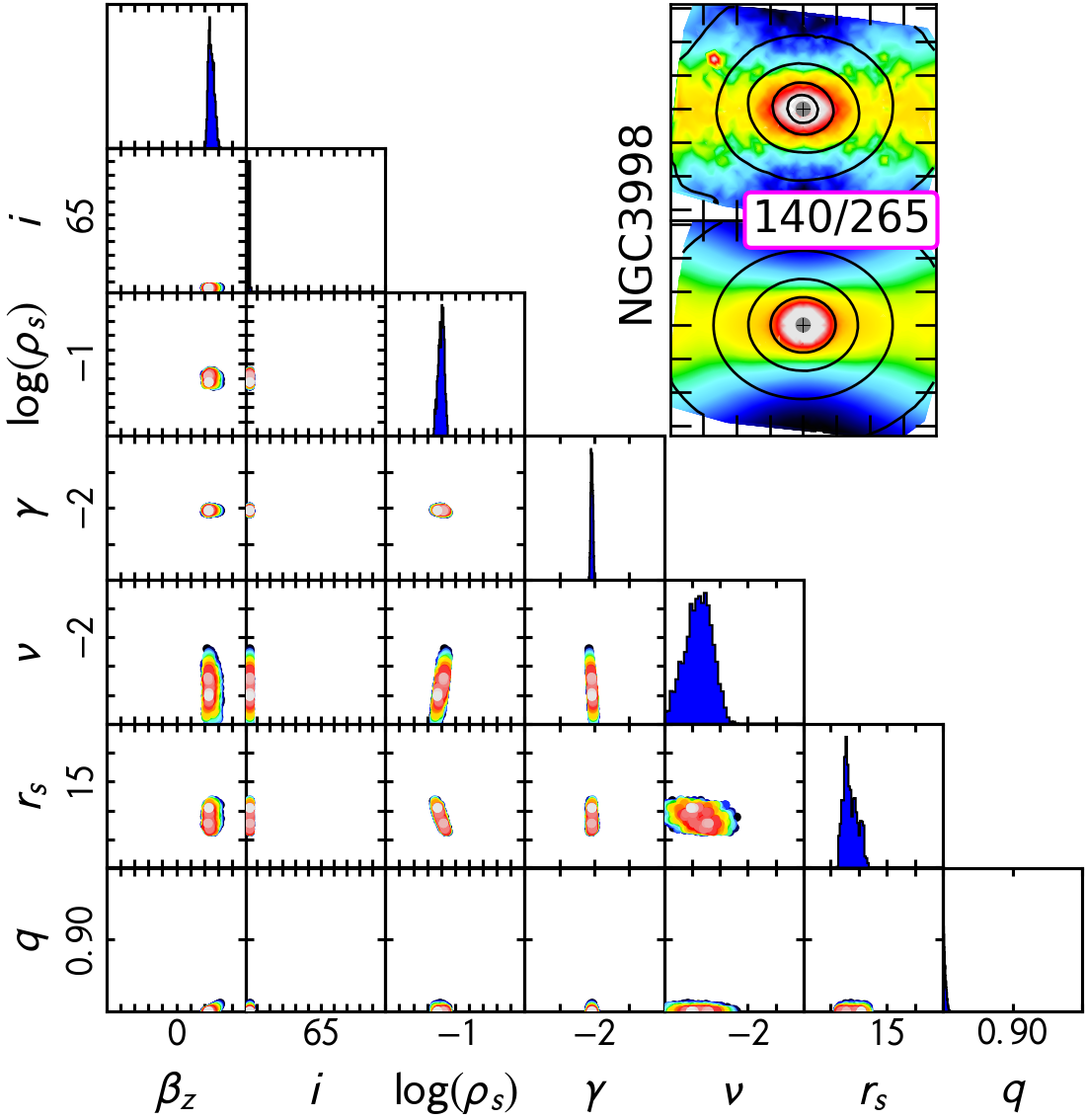}
	\hspace{5em}
	\includegraphics[resolution=150, height=0.311\textheight]{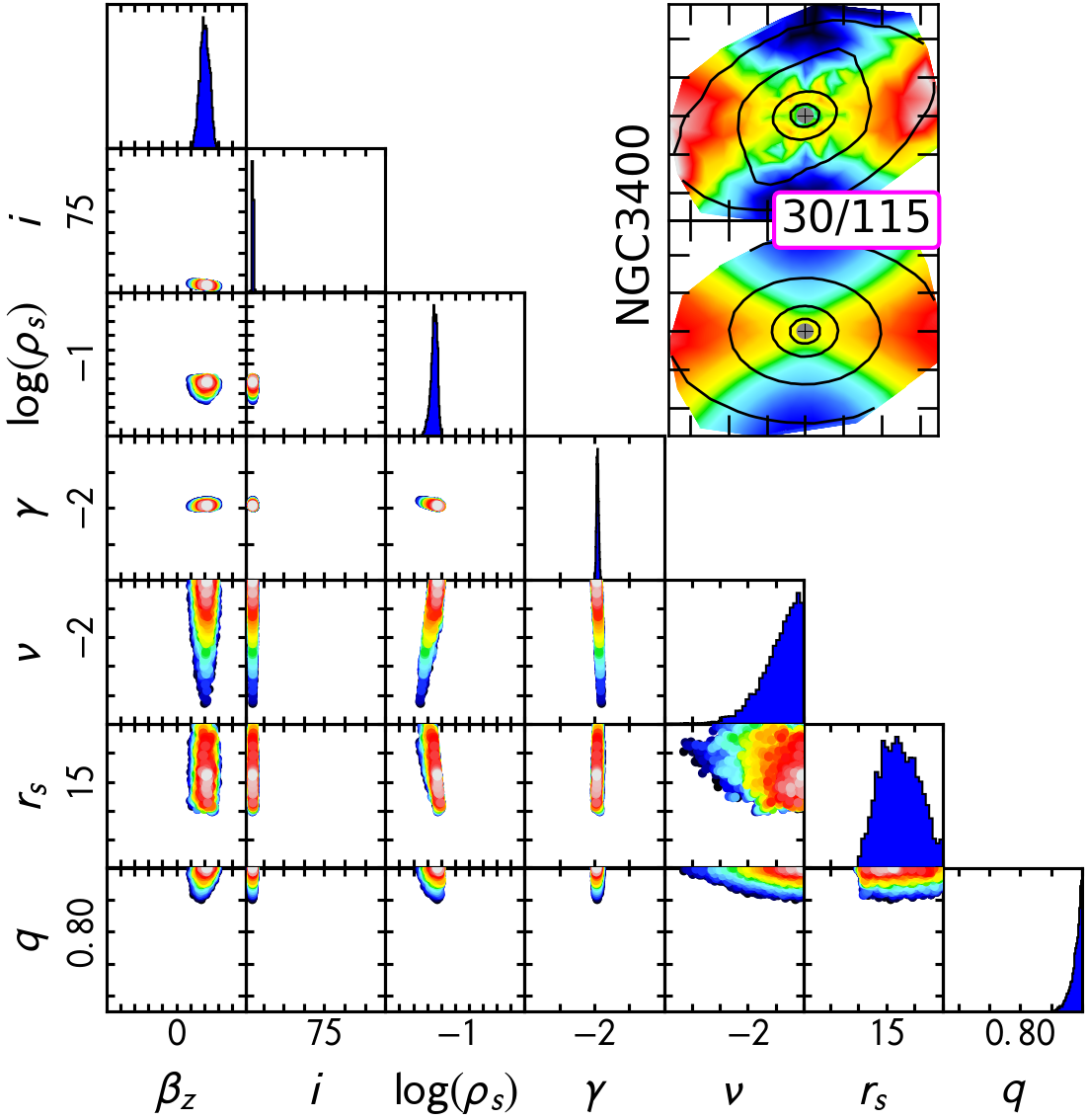}
}


%% file: append.tex
\begin{ThreePartTable}
	\begin{TableNotes}[para]
		\footnotesize
		Column (1) is the name of the galaxy. Columns (2), (3), and (4) are the mean logarithmic slopes, \(\gamma^\prime_{\mathrm{tot}} = \Delta\log_{10}\left(\rho_{\rm tot}\right) \big/ \Delta\log_{10}\left(r\right)\), of the total-mass density profile for \mi, \textbf{II}, and \textbf{III}, respectively, with typical measurement errors on these values of \(0.09\). Columns (5) and (6) are the mean logarithmic slope of the stellar-only mass density profile, applying the \ml{Salp} profiles, and assuming a constant \(M/L\), respectively. Columns (7) and (8) are the dark matter fractions for \mi\ and \textbf{II}, respectively. The corresponding dark matter fractions column for \miii\ was published as Column (3) of Table 1 in \protect\cite{cappellari13b}. An electronic version of this table is available from http://purl.org/atlas3d
	\end{TableNotes}
	\begin{longtable}{rd{2.2}d{2.2}d{2.2}d{2.2}d{2.2}d{2.2}d{2.2}}
			\caption[]{The total slopes and dark matter fractions for the models in this work.}
			\label{tab:data}\\
			\toprule
			\mc{Name} & \mc{\(\gamma^\prime_{\mathrm{tot, I}}\)} & \mc{\(\gamma^\prime_{\mathrm{tot, II}}\)} & \mc{\(\gamma^\prime_{\rm tot, III}\)} & \mc{\(\gamma^\prime_\mathrm{star}\)} & \mc{\(\gamma^\prime_\mathrm{star, III}\)} & \mc{\(f_{\mathrm{DM, I}}(r=R_e)\)} & \mc{\(f_{\mathrm{DM, II}}(r=R_e)\)}\\
		\mc{(1)} & \mc{(2)} & \mc{(3)} & \mc{(4)} & \mc{(5)} & \mc{(6)} & \mc{(7)} & \mc{(8)}\\\midrule
			\endfirsthead
			\multicolumn{8}{l}{\tablename\ \thetable\ -- \textit{continued}} \\
			\toprule
			\mc{Name} & \mc{\(\gamma^\prime_{\mathrm{tot, I}}\)} & \mc{\(\gamma^\prime_{\mathrm{tot, II}}\)} & \mc{\(\gamma^\prime_{\rm tot, III}\)} & \mc{\(\gamma^\prime_\mathrm{star}\)} & \mc{\(\gamma^\prime_\mathrm{star, III}\)} & \mc{\(f_{\mathrm{DM, I}}(r=R_e)\)} & \mc{\(f_{\mathrm{DM, II}}(r=R_e)\)}\\
		\mc{(1)} & \mc{(2)} & \mc{(3)} & \mc{(4)} & \mc{(5)} & \mc{(6)} & \mc{(7)} & \mc{(8)}\\\midrule
			\endhead
			\bottomrule
			\endfoot
			\insertTableNotes
			\endlastfoot
			    IC0560  &    -2.20     &    -2.28     &    -2.31     &    -2.37     &    -2.39    &    0.10     &     0.02   \\
			    IC0598  &    -1.92     &    -1.49     &    -2.02     &    -2.11     &    -2.27    &    0.20     &     0.71   \\
			    IC0676  &    -2.01     &    -2.24     &    -1.84     &    -1.98     &    -2.08    &    0.00     &     0.00   \\
			    IC0719  &    -1.69     &    -1.64     &    -1.78     &    -1.90     &    -1.82    &    0.31     &     0.11   \\
			    IC0782  &    -1.70     &    -1.74     &    -1.68     &    -2.23     &    -2.21    &    0.51     &     0.44   \\
			    IC1024  &    -1.81     &    -1.82     &    -1.33     &    -1.46     &    -1.61    &    0.18     &     0.31   \\
			    IC3631  &    -2.81     &    -3.05     &    -1.99     &    -1.83     &    -2.05    &    0.00     &     0.00   \\
			   NGC0448  &    -2.00     &    -2.00     &    -1.93     &    -2.26     &    -2.14    &    0.19     &     0.12   \\
			   NGC0474  &    -2.42     &    -2.81     &    -2.53     &    -2.51     &    -2.49    &    0.17     &     0.00   \\
			   NGC0502  &    -2.78     &    -3.04     &    -2.61     &    -2.37     &    -2.37    &    0.00     &     0.00   \\
			   NGC0509  &    -1.85     &    -1.62     &    -1.84     &    -1.98     &    -1.98    &    0.22     &     0.00   \\
			   NGC0516  &    -1.71     &    -1.56     &    -1.66     &    -1.85     &    -1.88    &    0.31     &     0.06   \\
			   NGC0524  &    -1.81     &    -1.81     &    -1.84     &    -2.29     &    -2.29    &    0.46     &     0.48   \\
			   NGC0525  &    -2.34     &    -2.55     &    -2.06     &    -2.39     &    -2.38    &    0.14     &     0.03   \\
			   NGC0661  &    -2.17     &    -2.16     &    -2.31     &    -2.52     &    -2.46    &    0.23     &     0.24   \\
			   NGC0680  &    -2.32     &    -2.44     &    -2.41     &    -2.63     &    -2.56    &    0.19     &     0.07   \\
			   NGC0770  &    -1.99     &    -1.88     &    -1.74     &    -2.23     &    -2.22    &    0.23     &     0.40   \\
			   NGC0821  &    -2.20     &    -2.27     &    -2.20     &    -2.48     &    -2.39    &    0.28     &     0.23   \\
			   NGC0936  &    -2.01     &    -1.97     &    -1.81     &    -2.35     &    -2.33    &    0.42     &     0.45   \\
			   NGC1023  &    -2.41     &    -2.78     &    -2.38     &    -2.40     &    -2.36    &    0.02     &     0.00   \\
			   NGC1121  &    -2.16     &    -2.06     &    -2.13     &    -2.60     &    -2.43    &    0.12     &     0.27   \\
			   NGC1222  &    -1.48     &    -1.30     &    -1.65     &    -2.53     &    -2.54    &    0.78     &     0.85   \\
			   NGC1248  &    -2.21     &    -2.46     &    -2.62     &    -2.37     &    -2.33    &    0.13     &     0.00   \\
			   NGC1266  &    -2.30     &    -2.12     &    -1.78     &    -1.78     &    -1.82    &    0.00     &     0.00   \\
			   NGC1289  &    -2.22     &    -2.30     &    -2.36     &    -2.59     &    -2.50    &    0.19     &     0.11   \\
			   NGC1665  &    -2.16     &    -2.04     &    -2.18     &    -2.19     &    -2.17    &    0.04     &     0.21   \\
			   NGC2481  &    -2.28     &    -2.28     &    -2.10     &    -2.67     &    -2.52    &    0.09     &     0.16   \\
			   NGC2549  &    -2.12     &    -2.12     &    -2.18     &    -2.47     &    -2.39    &    0.19     &     0.16   \\
			   NGC2577  &    -2.22     &    -2.29     &    -2.27     &    -2.44     &    -2.33    &    0.18     &     0.11   \\
			   NGC2592  &    -2.49     &    -2.43     &    -2.48     &    -2.58     &    -2.54    &    0.04     &     0.09   \\
			   NGC2594  &    -2.48     &    -2.46     &    -2.34     &    -2.81     &    -2.76    &    0.05     &     0.07   \\
			   NGC2679  &    -2.29     &    -2.72     &    -2.15     &    -2.30     &    -2.31    &    0.00     &     0.00   \\
			   NGC2685  &    -2.03     &    -2.04     &    -2.16     &    -2.35     &    -2.31    &    0.21     &     0.15   \\
			   NGC2695  &    -2.26     &    -2.32     &    -2.41     &    -2.55     &    -2.49    &    0.17     &     0.11   \\
			   NGC2698  &    -2.53     &    -2.81     &    -2.53     &    -2.76     &    -2.62    &    0.04     &     0.00   \\
			   NGC2699  &    -2.32     &    -2.31     &    -2.30     &    -2.54     &    -2.50    &    0.12     &     0.13   \\
			   NGC2764  &    -2.06     &    -2.02     &    -1.85     &    -1.73     &    -1.78    &    0.04     &     0.18   \\
			   NGC2768  &    -2.06     &    -2.57     &    -2.17     &    -2.23     &    -2.20    &    0.13     &     0.00   \\
			   NGC2778  &    -2.29     &    -2.28     &    -2.53     &    -2.56     &    -2.47    &    0.13     &     0.14   \\
			   NGC2824  &    -1.98     &    -1.98     &    -1.86     &    -2.45     &    -2.46    &    0.24     &     0.24   \\
			   NGC2852  &    -3.03     &    -3.11     &    -2.58     &    -2.62     &    -2.67    &    0.00     &     0.00   \\
			   NGC2859  &    -2.31     &    -2.66     &    -2.31     &    -2.51     &    -2.46    &    0.26     &     0.00   \\
			   NGC2880  &    -2.09     &    -2.27     &    -2.18     &    -2.46     &    -2.42    &    0.31     &     0.11   \\
			   NGC2950  &    -2.32     &    -2.31     &    -2.41     &    -2.63     &    -2.56    &    0.17     &     0.18   \\
			   NGC2962  &    -2.57     &    -3.42     &    -2.34     &    -2.47     &    -2.38    &    0.00     &     0.00   \\
			   NGC2974  &    -2.20     &    -2.26     &    -2.37     &    -2.44     &    -2.36    &    0.20     &     0.12   \\
			   NGC3032  &    -2.27     &    -2.12     &    -2.04     &    -2.52     &    -2.55    &    0.18     &     0.33   \\
			   NGC3073  &    -2.17     &    -2.51     &    -1.92     &    -2.59     &    -2.36    &    0.34     &     0.04   \\
			   NGC3098  &    -1.89     &    -1.84     &    -1.75     &    -2.22     &    -2.15    &    0.21     &     0.09   \\
			   NGC3156  &    -1.81     &    -1.77     &    -1.86     &    -1.77     &    -2.04    &    0.17     &     0.24   \\
			   NGC3182  &    -1.79     &    -1.34     &    -1.75     &    -2.36     &    -2.19    &    0.56     &     0.83   \\
			   NGC3193  &    -2.24     &    -2.40     &    -2.29     &    -2.38     &    -2.37    &    0.16     &     0.03   \\
			   NGC3226  &    -2.28     &    -2.29     &    -2.21     &    -2.22     &    -2.15    &    0.00     &     0.00   \\
			   NGC3230  &    -2.19     &    -2.08     &    -2.30     &    -2.44     &    -2.31    &    0.20     &     0.37   \\
			   NGC3245  &    -2.27     &    -1.99     &    -2.38     &    -2.55     &    -2.47    &    0.18     &     0.45   \\
			   NGC3248  &    -2.21     &    -2.35     &    -2.40     &    -2.54     &    -2.53    &    0.20     &     0.05   \\
			   NGC3301  &    -2.13     &    -2.06     &    -2.22     &    -2.36     &    -2.45    &    0.15     &     0.26   \\
			   NGC3377  &    -2.22     &    -2.10     &    -2.30     &    -2.53     &    -2.35    &    0.25     &     0.34   \\
			   NGC3379  &    -2.08     &    -2.14     &    -2.07     &    -2.44     &    -2.43    &    0.39     &     0.39   \\
			   NGC3384  &    -2.19     &    -2.28     &    -2.44     &    -2.73     &    -2.69    &    0.40     &     0.36   \\
			   NGC3400  &    -1.88     &    -1.85     &    -2.05     &    -2.15     &    -2.14    &    0.27     &     0.33   \\
			   NGC3412  &    -2.03     &    -1.95     &    -1.98     &    -2.55     &    -2.50    &    0.47     &     0.51   \\
			   NGC3414  &    -2.26     &    -2.18     &    -2.26     &    -2.47     &    -2.42    &    0.19     &     0.23   \\
			   NGC3457  &    -2.14     &    -2.14     &    -2.11     &    -2.31     &    -2.31    &    0.07     &     0.09   \\
			   NGC3458  &    -2.33     &    -2.37     &    -2.42     &    -2.75     &    -2.74    &    0.10     &     0.05   \\
			   NGC3489  &    -2.06     &    -2.08     &    -2.24     &    -2.38     &    -2.46    &    0.26     &     0.22   \\
			   NGC3499  &    -1.91     &    -2.01     &    -1.69     &    -2.67     &    -2.64    &    0.30     &     0.28   \\
			   NGC3522  &    -2.00     &    -2.02     &    -2.17     &    -2.53     &    -2.45    &    0.31     &     0.26   \\
			   NGC3530  &    -1.90     &    -1.86     &    -1.82     &    -2.45     &    -2.28    &    0.20     &     0.39   \\
			   NGC3595  &    -2.38     &    -2.15     &    -2.46     &    -2.59     &    -2.46    &    0.11     &     0.30   \\
			   NGC3599  &    -2.33     &    -2.58     &    -2.32     &    -2.42     &    -2.42    &    0.01     &     0.00   \\
			   NGC3605  &    -2.45     &    -2.49     &    -2.36     &    -2.39     &    -2.40    &    0.01     &     0.00   \\
			   NGC3607  &    -2.23     &    -2.44     &    -2.24     &    -2.44     &    -2.41    &    0.23     &     0.09   \\
			   NGC3608  &    -2.14     &    -2.14     &    -2.25     &    -2.41     &    -2.34    &    0.30     &     0.30   \\
			   NGC3610  &    -2.27     &    -2.27     &    -2.29     &    -2.53     &    -2.47    &    0.17     &     0.12   \\
			   NGC3613  &    -2.21     &    -2.38     &    -2.23     &    -2.39     &    -2.29    &    0.15     &     0.00   \\
			   NGC3619  &    -2.29     &    -2.22     &    -2.22     &    -2.56     &    -2.55    &    0.28     &     0.30   \\
			   NGC3626  &    -2.20     &    -2.16     &    -2.26     &    -2.38     &    -2.47    &    0.24     &     0.28   \\
			   NGC3630  &    -2.19     &    -2.19     &    -2.23     &    -2.82     &    -2.71    &    0.19     &     0.22   \\
			   NGC3640  &    -2.11     &    -2.21     &    -2.18     &    -2.37     &    -2.32    &    0.21     &     0.19   \\
			   NGC3641  &    -2.18     &    -2.04     &    -2.20     &    -3.05     &    -3.03    &    0.39     &     0.43   \\
			   NGC3648  &    -2.07     &    -2.05     &    -2.27     &    -2.69     &    -2.54    &    0.27     &     0.33   \\
			   NGC3658  &    -2.45     &    -2.57     &    -2.75     &    -2.54     &    -2.54    &    0.06     &     0.00   \\
			   NGC3665  &    -2.11     &    -2.24     &    -2.18     &    -2.29     &    -2.26    &    0.18     &     0.05   \\
			   NGC3674  &    -2.31     &    -2.38     &    -2.23     &    -2.68     &    -2.55    &    0.12     &     0.06   \\
			   NGC3694  &    -1.57     &    -1.52     &    -1.40     &    -2.35     &    -2.35    &    0.59     &     0.73   \\
			   NGC3757  &    -2.26     &    -2.23     &    -2.08     &    -2.79     &    -2.77    &    0.14     &     0.16   \\
			   NGC3796  &    -1.93     &    -1.91     &    -2.05     &    -2.42     &    -2.52    &    0.28     &     0.31   \\
			   NGC3838  &    -2.36     &    -2.46     &    -2.33     &    -2.88     &    -2.68    &    0.09     &     0.00   \\
			   NGC3941  &    -2.33     &    -2.37     &    -2.49     &    -2.48     &    -2.40    &    0.11     &     0.08   \\
			   NGC3945  &    -2.21     &    -2.49     &    -2.06     &    -2.46     &    -2.42    &    0.23     &     0.00   \\
			   NGC3998  &    -2.25     &    -2.23     &    -2.35     &    -2.48     &    -2.50    &    0.27     &     0.29   \\
			   NGC4026  &    -2.28     &    -2.25     &    -2.15     &    -2.32     &    -2.29    &    0.12     &     0.17   \\
			   NGC4036  &    -2.21     &    -2.41     &    -2.17     &    -2.27     &    -2.16    &    0.10     &     0.00   \\
			   NGC4078  &    -2.26     &    -2.33     &    -2.00     &    -2.44     &    -2.31    &    0.10     &     0.01   \\
			   NGC4111  &    -2.38     &    -2.38     &    -2.22     &    -1.71     &    -1.74    &    0.04     &     0.01   \\
			   NGC4119  &    -1.73     &    -1.89     &    -1.92     &    -1.90     &    -1.99    &    0.29     &     0.13   \\
			   NGC4143  &    -2.18     &    -2.09     &    -2.28     &    -2.45     &    -2.41    &    0.20     &     0.33   \\
			   NGC4150  &    -2.15     &    -2.12     &    -2.33     &    -2.23     &    -2.49    &    0.02     &     0.08   \\
			   NGC4168  &    -1.79     &    -1.80     &    -2.00     &    -2.19     &    -2.15    &    0.49     &     0.55   \\
			   NGC4179  &    -2.20     &    -2.20     &    -2.18     &    -2.41     &    -2.33    &    0.15     &     0.13   \\
			   NGC4191  &    -2.14     &    -2.18     &    -2.12     &    -2.46     &    -2.38    &    0.19     &     0.12   \\
			   NGC4203  &    -2.66     &    -2.78     &    -2.52     &    -2.60     &    -2.57    &    0.00     &     0.00   \\
			   NGC4215  &    -2.24     &    -2.27     &    -2.24     &    -2.46     &    -2.37    &    0.08     &     0.04   \\
			   NGC4233  &    -2.12     &    -2.13     &    -2.22     &    -2.62     &    -2.57    &    0.31     &     0.30   \\
			   NGC4249  &    -1.52     &    -1.46     &    -1.49     &    -2.35     &    -2.35    &    0.75     &     0.73   \\
			   NGC4251  &    -2.10     &    -2.09     &    -2.13     &    -2.28     &    -2.25    &    0.20     &     0.16   \\
			   NGC4255  &    -2.31     &    -2.42     &    -2.38     &    -2.79     &    -2.67    &    0.15     &     0.02   \\
			   NGC4259  &    -1.90     &    -1.80     &    -1.78     &    -2.53     &    -2.44    &    0.31     &     0.49   \\
			   NGC4261  &    -2.02     &    -1.96     &    -2.17     &    -2.39     &    -2.38    &    0.40     &     0.38   \\
			   NGC4262  &    -2.60     &    -2.56     &    -2.40     &    -2.82     &    -2.77    &    0.05     &     0.05   \\
			   NGC4264  &    -1.84     &    -1.88     &    -1.78     &    -2.19     &    -2.13    &    0.38     &     0.30   \\
			   NGC4267  &    -2.30     &    -2.27     &    -2.54     &    -2.65     &    -2.65    &    0.37     &     0.32   \\
			   NGC4268  &    -2.07     &    -1.90     &    -2.08     &    -2.52     &    -2.41    &    0.27     &     0.45   \\
			   NGC4270  &    -2.47     &    -2.84     &    -2.17     &    -2.30     &    -2.23    &    0.00     &     0.00   \\
			   NGC4278  &    -2.17     &    -2.23     &    -2.07     &    -2.41     &    -2.45    &    0.22     &     0.18   \\
			   NGC4281  &    -2.25     &    -2.40     &    -2.33     &    -2.37     &    -2.32    &    0.10     &     0.00   \\
			   NGC4283  &    -2.71     &    -2.74     &    -2.21     &    -2.45     &    -2.44    &    0.00     &     0.00   \\
			   NGC4324  &    -1.87     &    -1.85     &    -1.97     &    -2.28     &    -2.21    &    0.37     &     0.41   \\
			   NGC4339  &    -2.49     &    -2.88     &    -2.37     &    -2.30     &    -2.34    &    0.00     &     0.00   \\
			   NGC4340  &    -2.10     &    -2.18     &    -2.07     &    -2.40     &    -2.40    &    0.45     &     0.41   \\
			   NGC4342  &    -2.50     &    -2.49     &    -2.26     &    -2.65     &    -2.56    &    0.03     &     0.02   \\
			   NGC4346  &    -2.11     &    -2.09     &    -2.25     &    -2.58     &    -2.54    &    0.28     &     0.32   \\
			   NGC4350  &    -2.22     &    -2.24     &    -2.12     &    -2.46     &    -2.35    &    0.11     &     0.08   \\
			   NGC4365  &    -2.19     &    -2.11     &    -2.22     &    -2.41     &    -2.36    &    0.15     &     0.19   \\
			   NGC4371  &    -1.93     &    -1.95     &    -1.97     &    -2.17     &    -2.15    &    0.26     &     0.33   \\
			   NGC4374  &    -2.13     &    -2.17     &    -2.28     &    -2.37     &    -2.37    &    0.20     &     0.21   \\
			   NGC4377  &    -2.04     &    -1.96     &    -1.85     &    -2.60     &    -2.59    &    0.34     &     0.45   \\
			   NGC4379  &    -2.27     &    -2.27     &    -2.28     &    -2.31     &    -2.28    &    0.09     &     0.09   \\
			   NGC4382  &    -1.90     &    -2.07     &    -1.79     &    -2.09     &    -2.24    &    0.68     &     0.97   \\
			   NGC4387  &    -2.26     &    -2.28     &    -2.04     &    -2.14     &    -2.10    &    0.07     &     0.00   \\
			   NGC4406  &    -2.08     &    -1.98     &    -2.20     &    -2.22     &    -2.20    &    0.29     &     0.19   \\
			   NGC4417  &    -2.17     &    -2.13     &    -2.20     &    -2.40     &    -2.38    &    0.17     &     0.25   \\
			   NGC4425  &    -1.91     &    -1.92     &    -1.84     &    -2.05     &    -2.01    &    0.21     &     0.06   \\
			   NGC4429  &    -2.39     &    -2.99     &    -2.19     &    -2.20     &    -2.17    &    0.00     &     0.00   \\
			   NGC4434  &    -2.76     &    -2.95     &    -2.42     &    -2.41     &    -2.46    &    0.00     &     0.00   \\
			   NGC4435  &    -2.12     &    -2.15     &    -2.16     &    -2.27     &    -2.25    &    0.21     &     0.18   \\
			   NGC4442  &    -2.36     &    -2.75     &    -2.30     &    -2.33     &    -2.27    &    0.03     &     0.00   \\
			   NGC4452  &    -1.83     &    -1.63     &    -1.36     &    -1.56     &    -1.50    &    0.17     &     0.08   \\
			   NGC4458  &    -2.19     &    -2.08     &    -2.03     &    -2.26     &    -2.24    &    0.07     &     0.18   \\
			   NGC4459  &    -2.14     &    -2.10     &    -2.12     &    -2.34     &    -2.32    &    0.18     &     0.24   \\
			   NGC4461  &    -2.24     &    -2.31     &    -2.42     &    -2.51     &    -2.44    &    0.18     &     0.10   \\
			   NGC4472  &    -2.12     &    -2.36     &    -2.37     &    -2.24     &    -2.21    &    0.00     &     0.00   \\
			   NGC4473  &    -2.09     &    -2.11     &    -2.17     &    -2.56     &    -2.48    &    0.33     &     0.29   \\
			   NGC4474  &    -2.17     &    -2.45     &    -2.20     &    -2.48     &    -2.35    &    0.22     &     0.00   \\
			   NGC4476  &    -1.97     &    -2.03     &    -1.94     &    -2.29     &    -2.30    &    0.31     &     0.07   \\
			   NGC4477  &    -1.85     &    -1.70     &    -1.92     &    -2.42     &    -2.41    &    0.66     &     0.68   \\
			   NGC4478  &    -2.30     &    -2.28     &    -2.03     &    -2.00     &    -1.98    &    0.02     &     0.00   \\
			   NGC4483  &    -1.93     &    -1.92     &    -1.97     &    -2.30     &    -2.24    &    0.34     &     0.37   \\
			   NGC4486  &    -1.99     &    -1.89     &    -2.11     &    -2.16     &    -2.15    &    0.00     &     0.00   \\
			  NGC4486A  &    -2.81     &    -2.86     &    -2.20     &    -2.40     &    -2.41    &    0.00     &     0.00   \\
			   NGC4489  &    -1.66     &    -1.67     &    -1.48     &    -2.27     &    -2.27    &    0.68     &     0.61   \\
			   NGC4494  &    -2.00     &    -2.15     &    -2.13     &    -2.31     &    -2.26    &    0.38     &     0.40   \\
			   NGC4503  &    -2.27     &    -2.38     &    -2.38     &    -2.43     &    -2.34    &    0.13     &     0.02   \\
			   NGC4521  &    -1.97     &    -1.71     &    -1.90     &    -2.28     &    -2.21    &    0.29     &     0.66   \\
			   NGC4526  &    -2.26     &    -2.54     &    -2.24     &    -2.24     &    -2.24    &    0.00     &     0.00   \\
			   NGC4528  &    -2.15     &    -2.42     &    -2.40     &    -2.42     &    -2.38    &    0.17     &     0.00   \\
			   NGC4546  &    -2.33     &    -2.45     &    -2.43     &    -2.50     &    -2.41    &    0.10     &     0.00   \\
			   NGC4550  &    -1.79     &    -1.62     &    -1.60     &    -1.85     &    -1.83    &    0.28     &     0.05   \\
			   NGC4551  &    -2.12     &    -2.16     &    -2.06     &    -2.10     &    -2.06    &    0.08     &     0.00   \\
			   NGC4552  &    -2.07     &    -2.29     &    -2.20     &    -2.40     &    -2.38    &    0.42     &     0.28   \\
			   NGC4564  &    -2.27     &    -2.30     &    -2.28     &    -2.48     &    -2.36    &    0.11     &     0.07   \\
			   NGC4570  &    -2.14     &    -2.14     &    -2.13     &    -2.47     &    -2.38    &    0.17     &     0.24   \\
			   NGC4578  &    -2.04     &    -1.92     &    -2.05     &    -2.60     &    -2.46    &    0.48     &     0.55   \\
			   NGC4596  &    -2.01     &    -2.25     &    -2.27     &    -2.38     &    -2.35    &    0.34     &     0.22   \\
			   NGC4608  &    -2.05     &    -2.29     &    -3.12     &    -2.44     &    -2.44    &    0.31     &     0.00   \\
			   NGC4612  &    -1.81     &    -2.03     &    -1.75     &    -2.30     &    -2.38    &    0.54     &     0.33   \\
			   NGC4621  &    -2.22     &    -2.43     &    -2.29     &    -2.37     &    -2.30    &    0.14     &     0.00   \\
			   NGC4623  &    -1.61     &    -1.56     &    -1.64     &    -2.14     &    -2.05    &    0.52     &     0.73   \\
			   NGC4624  &    -1.99     &    -2.04     &    -2.15     &    -2.43     &    -2.42    &    0.43     &     0.43   \\
			   NGC4636  &    -1.90     &    -1.92     &    -1.97     &    -2.40     &    -2.34    &    0.56     &     0.64   \\
			   NGC4638  &    -2.15     &    -2.10     &    -2.07     &    -2.23     &    -2.12    &    0.12     &     0.03   \\
			   NGC4643  &    -2.15     &    -2.42     &    -2.19     &    -2.52     &    -2.49    &    0.26     &     0.03   \\
			   NGC4649  &    -2.09     &    -2.03     &    -2.11     &    -2.34     &    -2.34    &    0.04     &     0.04   \\
			   NGC4660  &    -2.25     &    -2.25     &    -2.27     &    -2.55     &    -2.45    &    0.11     &     0.09   \\
			   NGC4684  &    -1.95     &    -1.91     &    -1.90     &    -1.94     &    -1.98    &    0.14     &     0.11   \\
			   NGC4690  &    -2.30     &    -2.67     &    -2.31     &    -2.26     &    -2.25    &    0.00     &     0.00   \\
			   NGC4694  &    -2.65     &    -2.96     &    -2.22     &    -2.01     &    -2.13    &    0.00     &     0.00   \\
			   NGC4697  &    -2.10     &    -2.30     &    -2.18     &    -2.25     &    -2.19    &    0.19     &     0.11   \\
			   NGC4710  &    -2.19     &    -2.40     &    -1.65     &    -1.80     &    -1.80    &    0.00     &     0.00   \\
			   NGC4733  &    -2.01     &    -1.90     &    -1.97     &    -1.90     &    -1.90    &    0.00     &     0.00   \\
			   NGC4753  &    -2.02     &    -2.25     &    -1.99     &    -2.03     &    -2.05    &    0.00     &     0.00   \\
			   NGC4754  &    -2.11     &    -2.09     &    -2.22     &    -2.49     &    -2.44    &    0.29     &     0.31   \\
			   NGC4762  &    -2.06     &    -2.09     &    -1.98     &    -2.46     &    -2.32    &    0.24     &     0.22   \\
			   NGC4803  &    -1.50     &    -1.59     &    -1.52     &    -2.36     &    -2.33    &    0.63     &     0.68   \\
			   NGC5103  &    -2.07     &    -2.07     &    -2.08     &    -2.73     &    -2.53    &    0.21     &     0.26   \\
			   NGC5173  &    -2.02     &    -2.08     &    -1.90     &    -2.59     &    -2.59    &    0.31     &     0.25   \\
			   NGC5198  &    -2.31     &    -2.33     &    -2.32     &    -2.22     &    -2.19    &    0.00     &     0.00   \\
			   NGC5273  &    -1.83     &    -1.69     &    -1.57     &    -1.89     &    -2.07    &    0.16     &     0.27   \\
			   NGC5308  &    -2.20     &    -2.23     &    -2.12     &    -2.41     &    -2.33    &    0.11     &     0.01   \\
			   NGC5322  &    -2.16     &    -2.48     &    -2.28     &    -2.42     &    -2.39    &    0.27     &     0.05   \\
			   NGC5342  &    -2.04     &    -1.86     &    -2.03     &    -2.88     &    -2.61    &    0.22     &     0.49   \\
			   NGC5353  &    -1.99     &    -1.82     &    -1.84     &    -2.06     &    -2.00    &    0.24     &     0.11   \\
			   NGC5355  &    -1.67     &    -1.65     &    -1.73     &    -2.08     &    -2.24    &    0.44     &     0.59   \\
			   NGC5358  &    -1.80     &    -1.75     &    -1.86     &    -2.58     &    -2.44    &    0.40     &     0.52   \\
			   NGC5379  &    -1.18     &    -0.85     &    -1.40     &    -1.88     &    -1.89    &    0.94     &     0.71   \\
			   NGC5422  &    -2.11     &    -2.08     &    -2.13     &    -2.47     &    -2.41    &    0.25     &     0.35   \\
			   NGC5473  &    -1.95     &    -1.87     &    -1.94     &    -2.48     &    -2.44    &    0.52     &     0.57   \\
			   NGC5475  &    -1.86     &    -1.61     &    -1.91     &    -2.27     &    -2.31    &    0.27     &     0.82   \\
			   NGC5481  &    -2.49     &    -2.87     &    -2.50     &    -2.47     &    -2.44    &    0.00     &     0.00   \\
			   NGC5485  &    -1.96     &    -1.89     &    -2.20     &    -2.32     &    -2.27    &    0.33     &     0.35   \\
			   NGC5493  &    -2.06     &    -2.04     &    -2.16     &    -2.52     &    -2.32    &    0.24     &     0.27   \\
			   NGC5500  &    -2.19     &    -2.50     &    -2.15     &    -2.22     &    -2.27    &    0.04     &     0.00   \\
			   NGC5507  &    -2.15     &    -2.13     &    -2.26     &    -2.62     &    -2.55    &    0.23     &     0.27   \\
			   NGC5557  &    -2.29     &    -2.07     &    -2.35     &    -2.41     &    -2.36    &    0.15     &     0.26   \\
			   NGC5574  &    -2.11     &    -2.04     &    -1.97     &    -2.01     &    -2.02    &    0.08     &     0.02   \\
			   NGC5576  &    -2.53     &    -2.91     &    -2.58     &    -2.64     &    -2.55    &    0.07     &     0.00   \\
			   NGC5582  &    -2.12     &    -2.12     &    -2.07     &    -2.63     &    -2.51    &    0.50     &     0.51   \\
			   NGC5611  &    -2.28     &    -2.34     &    -2.21     &    -2.61     &    -2.45    &    0.10     &     0.03   \\
			   NGC5631  &    -2.14     &    -2.18     &    -2.31     &    -2.47     &    -2.42    &    0.30     &     0.28   \\
			   NGC5638  &    -1.77     &    -1.64     &    -2.29     &    -2.25     &    -2.24    &    0.55     &     0.62   \\
			   NGC5687  &    -2.24     &    -2.29     &    -2.33     &    -2.59     &    -2.49    &    0.29     &     0.26   \\
			   NGC5770  &    -2.65     &    -2.80     &    -2.38     &    -2.53     &    -2.53    &    0.00     &     0.00   \\
			   NGC5813  &    -2.03     &    -2.12     &    -1.91     &    -2.38     &    -2.31    &    0.62     &     0.63   \\
			   NGC5831  &    -2.21     &    -2.13     &    -2.25     &    -2.44     &    -2.37    &    0.18     &     0.23   \\
			   NGC5838  &    -2.46     &    -2.62     &    -2.52     &    -2.45     &    -2.42    &    0.07     &     0.00   \\
			   NGC5839  &    -2.42     &    -2.21     &    -2.49     &    -2.55     &    -2.51    &    0.10     &     0.28   \\
			   NGC5845  &    -2.73     &    -2.72     &    -2.01     &    -2.62     &    -2.57    &    0.02     &     0.01   \\
			   NGC5846  &    -1.91     &    -1.98     &    -2.14     &    -2.20     &    -2.20    &    0.30     &     0.24   \\
			   NGC5854  &    -2.06     &    -2.09     &    -2.14     &    -2.22     &    -2.22    &    0.18     &     0.09   \\
			   NGC5864  &    -2.13     &    -2.27     &    -2.02     &    -2.06     &    -1.96    &    0.08     &     0.00   \\
			   NGC5866  &    -1.92     &    -1.90     &    -1.93     &    -1.87     &    -1.96    &    0.00     &     0.00   \\
			   NGC5869  &    -2.17     &    -2.24     &    -2.18     &    -2.58     &    -2.47    &    0.29     &     0.24   \\
			   NGC6010  &    -2.14     &    -2.15     &    -2.21     &    -2.67     &    -2.54    &    0.18     &     0.17   \\
			   NGC6014  &    -1.80     &    -1.93     &    -1.61     &    -1.97     &    -1.95    &    0.21     &     0.00   \\
			   NGC6017  &    -2.58     &    -3.26     &    -2.23     &    -2.67     &    -2.57    &    0.03     &     0.00   \\
			   NGC6149  &    -1.98     &    -1.91     &    -1.92     &    -2.58     &    -2.52    &    0.28     &     0.44   \\
			   NGC6278  &    -2.32     &    -2.08     &    -2.33     &    -2.65     &    -2.63    &    0.12     &     0.37   \\
			   NGC6547  &    -2.21     &    -2.34     &    -2.29     &    -3.07     &    -2.67    &    0.23     &     0.04   \\
			   NGC6548  &    -1.64     &    -1.49     &    -2.40     &    -2.40     &    -2.41    &    0.84     &     0.85   \\
			   NGC6703  &    -2.40     &    -2.63     &    -2.49     &    -2.46     &    -2.46    &    0.00     &     0.00   \\
			   NGC6798  &    -2.08     &    -2.01     &    -2.04     &    -2.50     &    -2.40    &    0.23     &     0.38   \\
			   NGC7280  &    -2.12     &    -1.98     &    -2.03     &    -2.41     &    -2.47    &    0.20     &     0.38   \\
			   NGC7332  &    -2.25     &    -2.23     &    -2.26     &    -2.37     &    -2.42    &    0.07     &     0.16   \\
			   NGC7454  &    -2.01     &    -2.02     &    -2.06     &    -2.23     &    -2.18    &    0.20     &     0.19   \\
			   NGC7457  &    -1.65     &    -1.60     &    -1.71     &    -2.01     &    -2.06    &    0.51     &     0.51   \\
			   NGC7465  &    -2.16     &    -2.17     &    -1.98     &    -2.48     &    -2.46    &    0.16     &     0.06   \\
			   NGC7693  &    -1.96     &    -2.27     &    -2.44     &    -2.26     &    -2.29    &    0.27     &     0.00   \\
			   NGC7710  &    -2.31     &    -2.71     &    -1.81     &    -2.33     &    -2.32    &    0.09     &     0.00   \\
			 PGC016060  &    -1.91     &    -1.94     &    -1.97     &    -2.15     &    -2.08    &    0.20     &     0.02   \\
			 PGC028887  &    -2.04     &    -2.00     &    -2.06     &    -2.90     &    -2.75    &    0.32     &     0.37   \\
			 PGC029321  &    -1.37     &    -1.33     &    -1.15     &    -2.23     &    -2.14    &    0.72     &     0.70   \\
			 PGC035754  &    -2.58     &    -2.79     &    -2.49     &    -2.88     &    -2.75    &    0.08     &     0.02   \\
			 PGC042549  &    -2.05     &    -2.13     &    -2.23     &    -2.44     &    -2.40    &    0.20     &     0.08   \\
			 PGC044433  &    -2.11     &    -2.10     &    -2.07     &    -2.74     &    -2.69    &    0.16     &     0.07   \\
			 PGC050395  &    -1.43     &    -1.24     &    -1.52     &    -2.42     &    -2.44    &    0.72     &     0.67   \\
			 PGC051753  &    -1.79     &    -1.78     &    -1.65     &    -1.93     &    -1.84    &    0.31     &     0.22   \\
			 PGC054452  &    -1.94     &    -2.10     &    -1.89     &    -2.25     &    -2.24    &    0.27     &     0.09   \\
			 PGC056772  &    -2.22     &    -2.39     &    -2.14     &    -2.11     &    -2.06    &    0.13     &     0.00   \\
			 PGC061468  &    -1.33     &    -1.24     &    -1.12     &    -2.00     &    -2.01    &    0.73     &     0.65   \\
			 PGC170172  &    -1.94     &    -1.93     &    -2.13     &    -2.23     &    -2.33    &    0.22     &     0.25   \\
			  UGC03960  &    -2.22     &    -2.47     &    -2.35     &    -2.35     &    -2.43    &    0.20     &     0.00   \\
			  UGC04551  &    -2.43     &    -2.42     &    -1.92     &    -2.48     &    -2.46    &    0.09     &     0.08   \\
			  UGC05408  &    -3.29     &    -3.77     &    -2.50     &    -2.66     &    -2.71    &    0.00     &     0.00   \\
			  UGC06062  &    -2.00     &    -1.95     &    -1.93     &    -2.52     &    -2.43    &    0.29     &     0.39   \\
			  UGC06176  &    -1.80     &    -1.77     &    -1.73     &    -2.23     &    -2.22    &    0.37     &     0.46   \\
			  UGC08876  &    -2.46     &    -2.50     &    -2.36     &    -3.10     &    -2.89    &    0.06     &     0.02   \\
			  UGC09519  &    -2.04     &    -2.02     &    -2.15     &    -2.59     &    -2.66    &    0.17     &     0.10   \\
			\bottomrule
	\end{longtable}
\end{ThreePartTable}
